\documentclass[useAMS,usenatbib,twocolumn]{mnras}

\usepackage{amsmath}
\usepackage{graphicx}
\usepackage{dcolumn}
\usepackage{bm}
\usepackage{epsfig}
\usepackage{amssymb,latexsym,mathrsfs}
\usepackage{graphicx}
\usepackage{enumitem}
\usepackage{color}
\usepackage{hyperref}
\usepackage{subcaption}
\usepackage{float}
\usepackage{natbib}
\usepackage{tikz}
\usepackage{ulem}
\usetikzlibrary{positioning}
\usepackage[utf8]{inputenc}
\usepackage[T1]{fontenc}
\usepackage{geometry}
\usepackage{gensymb}
\usepackage{threeparttable}

\hypersetup{
    colorlinks=true,
    linkcolor=red,
    citecolor=blue,
} 

\def\apj{{ ApJ}}
\def\apjl{{ ApJL}}
\def\apjs{{ ApJS}}

\def\aap{{ A\&A}}

\def\mnras{{ MNRAS}}
\def\araa{{ ARA\&A}}

\def\nat {{ Nature}}

\newcommand{\be}{\begin{equation}}
\newcommand{\ee}{\end{equation}}
\newcommand{\bs}{\begin{split}} 
\newcommand{\bea}{\begin{eqnarray}}
\newcommand{\eea}{\end{eqnarray}}

\title[Nucleosynthesis and associated kilonovae]
{Nucleosynthesis in outflows of compact objects and detection prospects of associated kilonovae}

\author[Ekanger et al.]{
Nick Ekanger$^{1}$\thanks{enick1@vt.edu}, Mukul Bhattacharya$^{2}$, Shunsaku Horiuchi$^{1,3}$\\ 
${}^1$Center for Neutrino Physics, Department of Physics, Virginia Tech, Blacksburg, VA 24061, USA\\
${}^2$Department of Physics; Department of Astronomy \& Astrophysics; Center for Multimessenger Astrophysics, Institute for Gravitation\\
and the Cosmos, The Pennsylvania State University, University Park, PA 16802, USA \\
${}^3$Kavli IPMU (WPI), UTIAS, The University of Tokyo, Kashiwa, Chiba 277-8583, Japan 
} 

\begin{document}

\date{Accepted 2023 July 30; Received 2023 July 07; in original form 2023 March 08}

\pagerange{\pageref{firstpage}--\pageref{lastpage}} \pubyear{2023}

\maketitle

\label{firstpage}

\begin{abstract} 
We perform a comparative analysis of nucleosynthesis yields from binary neutron star (BNS) mergers, black hole-neutron star (BHNS) mergers, and core-collapse supernovae (CCSNe) with the goal of determining which are the most dominant sources of r-process enrichment observed in stars. We find that BNS and BHNS binaries may eject similar mass distributions of robust r-process nuclei post-merger (up to third peak and actinides, $A\sim200-240$), after accounting for the volumetric event rates. Magnetorotational (MR) CCSNe likely undergo a weak r-process (up to $A\sim140$) and contribute to the production of light element primary process (LEPP) nuclei, whereas typical thermal, neutrino-driven CCSNe only synthesize up to first r-process peak nuclei ($A\sim80-90$). We also find that the upper limit to the rate of MR CCSNe is $\lesssim1$ per cent the rate of typical thermal CCSNe; if the rate was higher, then weak r-process nuclei would be overproduced. Although the largest uncertainty is from the volumetric event rate, the prospects are encouraging for confirming these rates in the next few years with upcoming surveys. Using a simple model to estimate the resulting kilonova light curve from mergers and our set of fiducial merger parameters, we predict that $\sim7$ BNS and $\sim2$ BHNS events will be detectable per year by the Vera C. Rubin Observatory (LSST), with prior gravitational wave (GW) triggers.
\end{abstract}

\begin{keywords}
nuclear reactions, nucleosynthesis, abundances -- methods: numerical -- stars: winds, outflows
\end{keywords} 

\section{Introduction}\label{intro}

\begin{table*}
\caption{Range of typical values (and chosen fiducial values, in parentheses) of binary merger and core-collapse system parameters found in simulations. In general, the parameter ranges here are not uniformly distributed. `Dyn' refers to dynamical ejecta for the BNS and BHNS mergers, `wind' refers to viscous and neutrino-driven winds from BNS and BHNS mergers, `MR' refers to magnetorotational, and `therm' refers to typical thermal CCSNe. $M_{\rm ej}$ is the ejecta mass for each scenario, $r_0$ is the initial radius (roughly corresponds to where the outflow begins) used to calculate the initial density for {\tt SkyNet}, $\theta_{\rm ej}$ is the polar angle within which the material is ejected, and `Refs' refers to the simulation references listed below the table. The mass ejected for BNS and BHNS dynamical/wind ejecta are calculated using the equations (\ref{bnsdynamical}) - (\ref{bhnsremnant}), assuming ALF2 NS EOS. BHNS average ejecta mass is integrated over distributions of $M_{\rm BH}$ and $\chi_{\rm BH}$.}
\begin{threeparttable}
\centering
\def\arraystretch{1.4}
\begin{tabular}{|c|cccccc|c|}
\hline
\textbf{Scenario} & \boldmath{$S$} [\boldmath{${\rm k_B\ nuc^{-1}}$}] & \boldmath{$ \tau_{\rm exp}$} [\textbf{ms}] & \boldmath{$Y_e$} & \boldmath{$M_{\rm ej}$} [$10^{-2}~M_{\odot}$] & $\mathbf{r_0}$ [\textbf{km}] & \boldmath{$\theta_{\rm ej}$} [\boldmath{\degree}] & \textbf{Refs}\\
\hline
BNS (dyn)&5 - 40 (10)&10 - 20 (10)&0.01 - 0.3 (0.15)&0.02 - 5 (0.5)&0 - 500 (500)&10 - 60 (30)&1\\
BNS (wind)&20 (20)&30 (30)&0.2 - 0.4 (0.35)&0.1 - 5 (0.2)&$\sim500$ (500)&$\sim180$ (180)&\\
\hline
BHNS (dyn)&0.5 - 10 (10)&1 - 10 (10)&0.05 - 0.1 (0.1)&0.02 - 10 (2)&$\sim500$ (500)&$\sim30$ (30)&2\\
BHNS (wind)&10 - 100 (10)&10 - 100 (30)&0.1 - 0.5 (0.3)&0.7 - 8 (7)&200 - 1000 (500)&65 - 180 (180)&\\
\hline
CCSN (MR)&5 - 90 (20)&1 - 60 (10)&0.15 - 0.6 (0.35)&0.3 - 300 (5)&10 - 50 (30)&20 - 45 (30)&3\\
CCSN (therm)&50 - 250 (100)&10 - 500 (50)&0.4 - 0.6 (0.45)&0.07 - 200 (0.07)&10 - 50 (30)&180 (180)&4\\ \hline
\end{tabular}
\begin{tablenotes}
\item[1] BNS merger sources: \citet{Korobkin:2012uy}, \citet{2015just}, \citet{Lippuner:2015gwa}, \citet{Bovard:2017mvn}, \citet{Radice:2018pdn}, \citet{Zhu:2020eyk},  \citet{Chen:2022nsj}, \citet{combi2022}, \citet{deHaas2022}, 
\citet{fujibayashi2022}, \citet{Kiuchi:2022nin},
\citet{Kullmann:2021gvo},
\citet{rosswog2022}
\item[2] BHNS merger sources: \citet{Korobkin:2012uy}, \citet{2015just}, \citet{Roberts:2016igt}, \citet{Bhattacharya:2018lmw}, \citet{Fujibayashi:2020jfr}, \citet{Kyutoku:2021icp}, \citet{Hayashi:2021oxy}
\item[3] Magnetorotational CCSN sources: \citet{Winteler:2012hu}, \citet{Vlasov:2014ara}, \citet{Nishimura:2015nca}, \citet{Vlasov:2017nou}, \citet{Halevi:2018vgp}, \citet{Reichert:2020mjo}, \citet{Bhattacharya:2021cjc}, \citet{Desai:2022lyi},
\citet{Ekanger:2022tia}, 
\citet{reichert2022}
\item[4] Thermal CCSN sources: \citet{Takahashi:1994yz}, \citet{Qian:1996xt}, \citet{Goriely:2016gfe}, \citet{Bliss:2018nhk}, \citet{Witt:2021gwk}, \citet{Psaltis:2022jgr}
\end{tablenotes}
\end{threeparttable}\label{table:typicalparameterrange}
\end{table*}

Answering the open question of where the elements of the periodic table are sourced from is a fundamental undertaking in astrophysics and it has seen significant progress in recent years. The production sites for the heaviest of elements like europium, gold, and uranium are not yet fully characterized. These heavy nuclei, created through the rapid neutron-capture process (or r-process), require extremely dense, hot, and neutron-rich environments to be synthesized \citep{b2fh}, so compact object mergers and core-collapse supernovae (CCSNe) are attractive candidates {\it a priori} and are supported in literature (see e.g. \citealt{Kasen:2017sxr,Siegel:2018zxq,Bartos:2019cec,Yong:2021nkh,Bhattacharya:2021cjc,Ekanger:2022tia} for recent works, and \citealt{Thielemann:2017acv,Horowitz:2018ndv,Kajino:2019abv,Cowan:2019pkx,Arcones:2022jer} for reviews). In recent years, many have studied these candidates theoretically as promising sites of heavy element nucleosynthesis, especially through the combination of numerical methods (see Table~\ref{table:typicalparameterrange} in $\S$\ref{methods} for a representative list of references) and nuclear reaction networks like {\tt WinNet} \citep{2014PhDT.......206W}, {\tt PRISM} \citep{Mumpower_2017}, and {\tt SkyNet} \citep{Lippuner_2017}.

In parallel, another method used to understand this question involves considering the abundance evolution over time for all elements, known as galactic chemical evolution (GCE; see, e.g., \citealt{Cote:2016vla,Cote:2017evr,Hotokezaka:2018aui,Kobayashi:2020jes}). GCE studies combine galaxy evolution, population syntheses of transients like mergers and supernovae, and nucleosynthesis models to explain the sources of the periodic table (\citealt{Cote:2018qku,Kobayashi:2022qlk}). Considering a holistic view of the sources of all the elements, that can fold in nucleosynthesis results, is an important, and independent, theoretical approach for understanding the sources of the r-process elements. It is also important to look for observational signatures of the r-process having occurred. 

In concert with theory, astronomers can observe evidence of the robust r-process. Close to home, abundances of very heavy nuclei up to and including actinides have been measured (e.g., \citealt{Arnould:2007gh,asplundabundances}). The solar abundance distribution reveals a characteristic pattern of three peaks at mass numbers A $\sim82,~130,~{\rm and}~196$. This pattern is also seen in metal-poor stars (e.g., \citealt{Honda:2006kp,sneden2008ARA&A..46..241S,HD222925}). However, the situation is more nuanced in the case of metal-poor stars. Many of these stars might have only been enriched with heavy elements once or a few times \citep{Holmbeck:2022mog}, and can exhibit some variance in the abundance of some intermediate-mass elements which suggests multiple source classes (see e.g. \citealt{2012A&A...545A..31H,2017ApJ...837....8A,2018ApJ...864...43C,2018A&A...611A..30S}). Comparing stellar abundance data to nucleosynthesis results is an important task not just to confirm the validity of physical models and nucleosynthesis networks, but also to observe patterns in these observations (see, e.g., \citealt{2015A&A...579A...8W,Reichert:2020mjo,Roederer:2022exr}).

\begin{figure*}
\includegraphics[width=\linewidth]{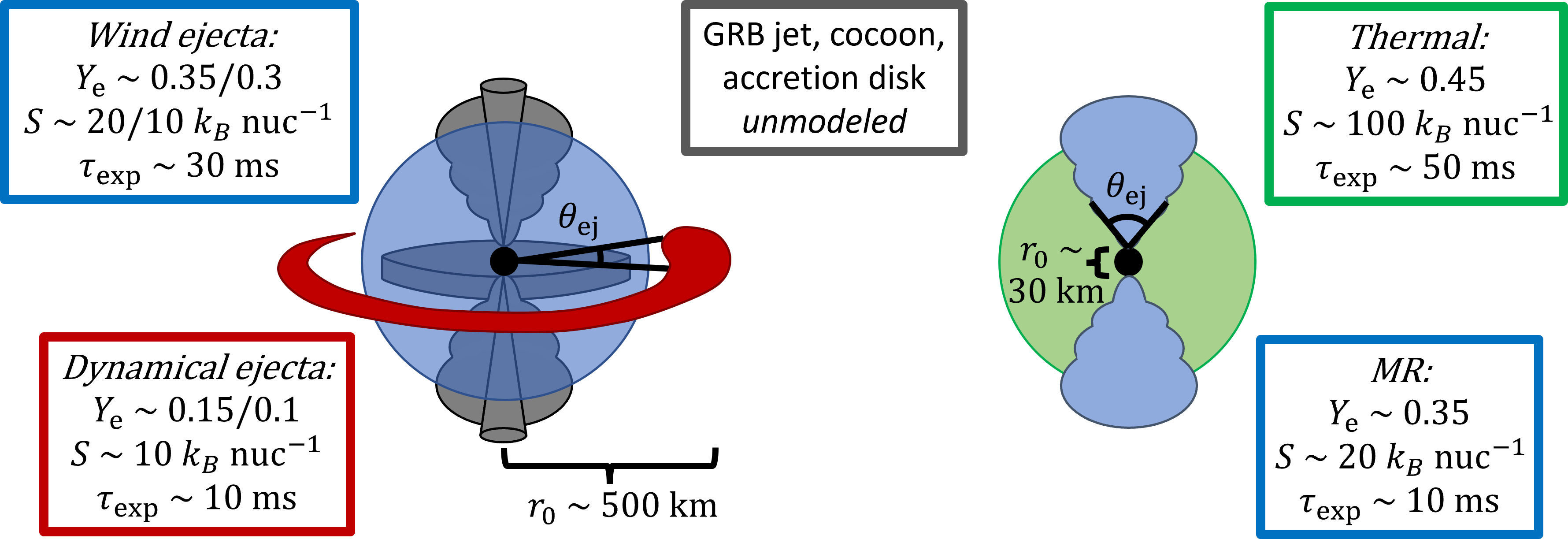}
\caption{Schematic diagram of the outflows arising from transients explored in this paper. We also show the fiducial parameters relevant for nucleosynthesis for each scenario we consider. The physical parameters are described in $\S$\ref{massejected} and $\S$\ref{networkandparameters} with a full description of their ranges and relevant sources are listed in Table~\ref{table:typicalparameterrange}. \textit{Left-hand panel:} Compact object mergers including BNS and BHNS. The blue component is the isotropic wind ejecta and the red component is the anisotropic dynamical ejecta. Gray components are unmodeled in our study but relevant for understanding the dynamics of these environments. \textit{Right-hand panel:} CCSNe including typical thermal, spherical outflows (in green) and magnetorotational, jetted outflows (in blue).}
\label{fig:schematic}
\end{figure*}

Alongside the measurement of stellar abundances, kilonovae - whose luminosity is derived from the decay products of unstable heavy nuclei synthesized through r-process (e.g., \citealt{Li:1998bw,Metzger:2010sy}) - are also detectable in ultraviolet, optical, and infrared. In 2017, the association of gravitational wave (GW) event GW170817 \citep{LIGOScientific:2017vwq} with a delayed short gamma-ray burst (GRB) (\citealt{LIGOScientific:2017zic,Goldstein:2017mmi,Savchenko:2017ffs}) and an electromagnetic signal AT2017gfo (\citealt{LIGOScientific:2017ync,Coulter:2017wya,DES:2017kbs,Valenti:2017ngx}) confirmed binary neutron star (BNS) mergers as sites of r-process nucleosynthesis. Despite this being the only such event detected, compact object mergers, especially BNS mergers, are likely to be one of the main sources of r-process elements (\citealt{Hotokezaka:2015zea,Beniamini:2016hoq,Tarumi:2021xvw}). Kilonovae from other compact object mergers such as black hole-neutron star (BHNS) mergers have not been observed yet, so the jury is still out on whether there are additional r-process nucleosynthesis sites and the relative contribution these make to the observed stellar abundances (see also \citealt{ji2019ApJ...882...40J}). 

To that end, we analyze and compare the nucleosynthesis yields of several astrophysical sites to help determine the relative contributions of r-process enrichment, with an emphasis on BHNS mergers which have not been as well-studied compared to BNS mergers. Figure \ref{fig:schematic} shows how we model compact object mergers with two components - the wind and dynamical ejecta - and differentiate them from CCSNe. The landscape for astrophysical phenomena associated with CCSNe, like protomagnetar outflows (\citealt{1992ApJ...392L...9D,1992Natur.357..472U,Metzger:2010pp,Vlasov:2014ara,Vlasov:2017nou,Bhattacharya:2021cjc,Ekanger:2022tia}), long-duration GRBs (\citealt{Wheeler_2000,Thompson_2004}), and collapsar accretion discs (\citealt{MacFadyen:1998vz,Pruet:2003ts,Kohri:2005tq,Siegel:2018zxq}), is broad. We focus on the typical thermal CCSNe for which winds are driven by neutrinos and magnetorotational (MR) CCSNe where winds are driven by the protoneutron star's magnetic field and rotation rate. This latter case of MR CCSNe are intriguing as potential sites for heavy element nucleosynthesis in theory as well as in simulations (see the references in Table~\ref{table:typicalparameterrange}), but the observational evidence is somewhat limited at present (see e.g. \citealt{Yong:2021nkh}). Figure~\ref{fig:schematic} also shows the initial parameters required to estimate the nucleosynthesis yields (in the left-hand panel, the parameters are listed in the order of BNS/BHNS, see also Table~\ref{table:typicalparameterrange} in $\S$\ref{methods}). We find that the relative contribution of BNS and BHNS mergers are similar, and they both synthesize third r-process peak nuclei and actinides. Their volumetric event rates, however, are poorly constrained. We also find that typical thermal CCSNe synthesize first peak and LEPP nuclei, while MR CCSNe likely synthesize up to second r-process peak nuclei. Finally, we show that the detection prospects for kilonova resulting from compact object mergers, associated with prior GW triggers, is promising in the next few years. This will also shed light on the dynamics and rates of compact mergers. 

This paper is organized as follows. In $\S$\ref{methods}, we describe the methods used, including using analytical fits to numerical relativity simulation results to estimate mass ejected in compact object mergers, a description of the nuclear reaction network used in this study, and how we parametrize the properties of these astrophysical systems. In $\S$\ref{results}, we discuss the nucleosynthesis results from each of our merger and core-collapse scenarios, and compare them with the measured stellar abundances of heavy nuclei from the Sun and a metal-poor star. In $\S$\ref{prospects}, we use a simple model to estimate the kilonova emission expected from the decay of unstable r-process nuclei and prospects for detecting more kilonovae within the next few years with LSST. In $\S$\ref{discussion} we discuss the relative importance of uncertainties for this paper and assumptions involved in our models. Finally, we summarize our results and conclude in $\S$\ref{summary}. 

\section{Methods}\label{methods}

Here we describe the methods used to estimate the abundance of heavy elements synthesized in BNS mergers, BHNS mergers, and CCSNe. We rely on several recent numerical simulations for each of these scenarios to guide our estimates for the mass ejected (post-merger or core collapse), entropy, expansion time-scale, and electron fraction which are all inputs to the nuclear reaction network used to compute nucleosynthesis yields.

\subsection{Estimating mass ejected}\label{massejected}

We estimate the mass ejected in the two merger scenarios from analytical expressions inferred from recent numerical simulations. For the BNS scenario, we want to estimate the dynamical ejecta and viscous/neutrino-driven wind (also known as secular) mass. BNS dynamical ejecta mass is estimated from \citep{Kruger:2020gig}
\begin{equation}\label{bnsdynamical}
\frac{M_{\rm dyn}^{\rm BNS}}{10^{-3}M_{\odot}}=\left(\frac{a}{C_1}+b\left(\frac{M_2}{M_1}\right)^n+cC_1\right)M_1+(1\leftrightarrow 2),
\end{equation}
where $a=-9.3335$, $b=114.17$, $c=-337.56$, and $n=1.5465$. $C_i=GM_i/(R_ic^2)$ is the compactness of an NS with gravitational mass $M_i$ and radius $R_i$. Negative values obtained from equation (\ref{bnsdynamical}) represent zero dynamical mass ejected post-BNS merger. 

The BNS wind ejecta mass is estimated as a fraction of the remnant disc mass $M_{\rm wind}^{\rm BNS}\approx M_{\rm disc}^{\rm BNS}\xi_{\rm wind}$ \citep{Kruger:2020gig}, where
\begin{equation}\label{bnsdisk}
    M_{\rm disc}^{\rm BNS}=M_1\max\left(5\times10^{-4},(fC_1+g)^h\right).
\end{equation}
Here, $f=-8.1324$, $g=1.482$ and $h=1.7784$, and the subscript `1' denotes the less massive of the two neutron stars. We consider $\xi_{\rm wind}\approx0.23$ after adding the contributions from neutrino-driven and viscous ejecta from \citet{Radice:2018pdn}: $\sim0.03\pm0.015$ for neutrino-driven component and $\sim0.2\pm0.1$ for the viscous ejecta. BNS wind ejecta is assumed to be approximately isotropic.

The mass of the BHNS dynamical ejecta is estimated as \citep{Kruger:2020gig}
\begin{equation}\label{bhnsdynejecta}
    \frac{M_{\rm dyn}^{\rm BHNS}}{M_{\rm NS}^b}=a_1Q^{n_1}\frac{1-2C_{\rm NS}}{C_{\rm NS}}-a_2Q^{n_2}\frac{R_{\rm ISCO}}{M_{\rm BH}}+a_4,
\end{equation}
where $a_1=0.007116$, $a_2=0.001436$, $a_4=-0.02762$, $n_1=0.8636$, and $n_2=1.6840$. $M_{\rm NS}^b$ is the NS baryonic mass, and $Q=M_{\rm BH}/M_{\rm NS}$ is the binary mass ratio. $R_{\rm ISCO}$ is given by \citep{Bardeen:1972fi}
\begin{equation}
    \frac{R_{\rm ISCO}}{M_{\rm BH}}=3+Z_2-{\rm sign}(\chi_{\rm BH})\sqrt{(3-Z_1)(3+Z_1+2Z_2)},
\end{equation}
where 
\begin{equation}
    Z_1=1+(1-\chi_{\rm BH}^2)^{1/3}\left[(1+\chi_{\rm BH})^{1/3}+(1-\chi_{\rm BH})^{1/3}\right],
\end{equation}
and
\begin{equation}
    Z_2=\sqrt{3\chi_{\rm BH}^2+Z_1^2},
\end{equation}
where $\chi_{\rm BH}$ is the dimensionless spin parameter of the BH. Depending on the initial BH and NS parameters, BHNS mergers may or may not have any dynamical mass ejected. If the tidal forces from the BH on the NS surface are not strong enough (i.e., $R_{\rm TD}\sim R_{\rm NS}(M_{\rm BH}/M_{\rm NS})^{1/3}~<R_{\rm ISCO}$ where $R_{\rm TD}$ is the tidal disruption radius), the NS remains intact as it reaches the innermost stable circular orbit radius, $R_{\rm ISCO}$, and plunges directly into the BH. This generally happens if the NS is very compact and/or the BH has a high mass, $M_{\rm BH}$, and low spin magnitude, $\chi_{\rm BH}$. However, if the tidal forces are strong enough, part of the disrupted NS material forms an accretion disc around the BH and the rest becomes (unbound) dynamical ejecta. In equation (\ref{bhnsdynejecta}), negative values again represent zero dynamical mass ejected post-BHNS merger.

The accretion disc surrounding the BH may give rise to matter outflows in addition to the dynamical ejecta. This BHNS wind ejecta is estimated from
\begin{equation}
    \frac{M_{\rm wind}^{\rm BHNS}}{M_{\rm disc}^{\rm BHNS}}=\xi_1+\frac{\xi_2-\xi_1}{1+e^{1.5(Q-3)}},
\end{equation}
where $\xi_1 \sim 0.04-0.32$ and $\xi_2 \sim 0.14-0.44$ \citep{Raaijmakers:2021slr}. For our analysis, we choose the central values $\xi_1=0.18$ and $\xi_2=0.29$ for these parameters. Unlike the dynamical ejecta, winds originating from the remnant accretion disc are spherically symmetric, as is also the case with disc winds from BNS merger remnants. 

The BHNS disc mass is given by 
\begin{equation}
    M_{\rm disc}^{\rm BHNS}=M_{\rm rem}^{\rm BHNS}-M_{\rm dyn}^{\rm BHNS},
\end{equation}
where 
\begin{equation}\label{bhnsremnant}
    M_{\rm rem}^{\rm BHNS}=M_{\rm NS}^b\left[\max\left(\alpha\frac{1-2C_{\rm NS}}{\eta^{1/3}}-\beta\frac{R_{\rm ISCO}}{M_{\rm BH}}\frac{C_{\rm NS}}{\eta}+\gamma,0\right)\right]^{\delta}
\end{equation}
is the remnant mass outside the BH $t \sim 10\, {\rm ms}$ after merger. Here, $\alpha=0.406$, $\beta=0.139$, $\gamma=0.255$, $\delta=1.761$, and $\eta=Q/(1+Q)^2$ are fit parameters obtained from numerical relativity simulations \citep{Foucart:2018rjc}. It should be noted that the mass and velocity (see equation \ref{velocityejecta}) estimates from these analytical fits are limited to the range of parameters: $Q =1-7$, $\chi_{\rm BH} = -0.5-0.97$, and $C_{\rm NS} = 0.13-0.18$. We use the ALF2 \citep{ALF2EOS} equation of state (EOS), which is consistent with the recent NICER and GW measurements (see e.g. fig.~11 from \citealt{Kyutoku:2021icp}). A discussion of how the NS EOS affects the kilonova light curves and detection rates is presented in Appendix~\ref{eosappendix}.

For both thermal and MR CCSNe, we look to simulations for reasonable values of mass ejected in the absence of fitting formulae. The mass ejected in CCSN outflows can also vary by several orders of magnitude (see Table~\ref{table:typicalparameterrange} and references listed within). In thermal CCSN outflows, we assume $M_{\rm ej}\sim7\times10^{-4}\,M_{\odot}$ so as not to overproduce r-process nuclei (\citealt{Takahashi:1994yz,Goriely:2016gfe}). In MR CCSN outflows, we assume a representative value of $M_{\rm ej}\sim5\times10^{-2}\,M_{\odot}$ which lies within the range of total mass ejected (see Table~\ref{table:typicalparameterrange} references).

\subsection{Nuclear reaction network and thermodynamic parameters}\label{networkandparameters}

We use the nuclear reaction network {\tt SkyNet} \citep{Lippuner_2017} to calculate the nucleosynthesis yields in our astrophysical sites of interest. In general, {\tt SkyNet} requires the matter density ($\rho$), temperature ($T$), and electron fraction ($Y_e$) as functions of time to calculate abundances of nuclei synthesized. We estimate the density evolution as (e.g., \citealt{Lippuner:2015gwa})
\begin{equation}\label{densityevo}
\begin{split}
\rho(t)=
\begin{cases} 
      \rho_0 e^{-t/\tau_{\rm exp}}, \hfill t<3\tau_{\rm exp}\\
      \rho_0 \left(\frac{3\tau_{\rm exp}}{et}\right)^3, \hfill t \geq 3\tau_{\rm exp}.
\end{cases}
\end{split}
\end{equation}
where $\tau_{\rm exp}$ is the expansion time-scale of the outflow. Although $\tau_{\rm exp}$ is defined differently in many works, generally it is the characteristic time-scale that signifies how fast the density of the outflow decreases (see e.g. \citealt{Lippuner:2015gwa}). This density evolution is motivated even for MR CCSNe with jetted outflows. Recently, \citet{Bhattacharya:2022btx} showed that magnetized jets remain uncollimated during their evolution, since they generally have a higher jet pressure and deposit more energy on to the cocoon, compared to hydrodynamical jets, before they break out. This is especially true for larger and less dense progenitors like red supergiants since they have longer breakout times compared to Wolf-Rayet stars. 

We estimate the initial outflow density from each component as $\rho_0=M_{\rm ej}/(2\pi r_0\int_0^{\theta_{\rm ej}}\sin\theta d\theta)$, where $r_0$ is the radius at which the initial temperature is calculated and approximately where nucleosynthesis starts and $\theta_{\rm ej}$ is the opening angle of the outflow. 
We estimate the initial temperature by using nuclear statistical equilibrium (NSE), given the initial density, entropy, and electron fraction. We then let the network evolve via the self-heating mode, where the temperature evolution is self-consistent and considers heating from the nuclear reactions. For our forward reaction rates, we use the REACLIB reaction database of \citet{reaclib} and use detailed balance to calculate inverse reactions so that it is consistent with NSE. We use 7836 species up to $Z=112$ and $A=337$. Lastly, we evolve {\tt SkyNet} until $10^9\,{\rm s}$ such that the majority of unstable nuclei have decayed.

The initial conditions required to fully characterize the nucleosynthesis (entropy $S$, expansion time-scale $\tau_{\rm exp}$, electron fraction $Y_e$, mass ejected $M_{\rm ej}$, initial radius $r_0$, and opening angle $\theta_{\rm ej}$) can be inferred from recent simulations of these systems and related literature. In Table \ref{table:typicalparameterrange}, we show reasonable ranges for the parameters of interest based on the references listed within. We also show the most `typical' values within these ranges, in parentheses, which we use for analysis in the rest of this paper unless stated otherwise. These represent the most probable values based on simulations. In reality, outflows from merger and CCSNe scenarios are complex regions where the thermodynamic and nuclear properties vary, with location in the outflow and as a function of time. Our choices here, however, serve to capture the average nucleosynthesis that one could expect to see from each scenario.

\begin{figure*}
\includegraphics[width=\linewidth]{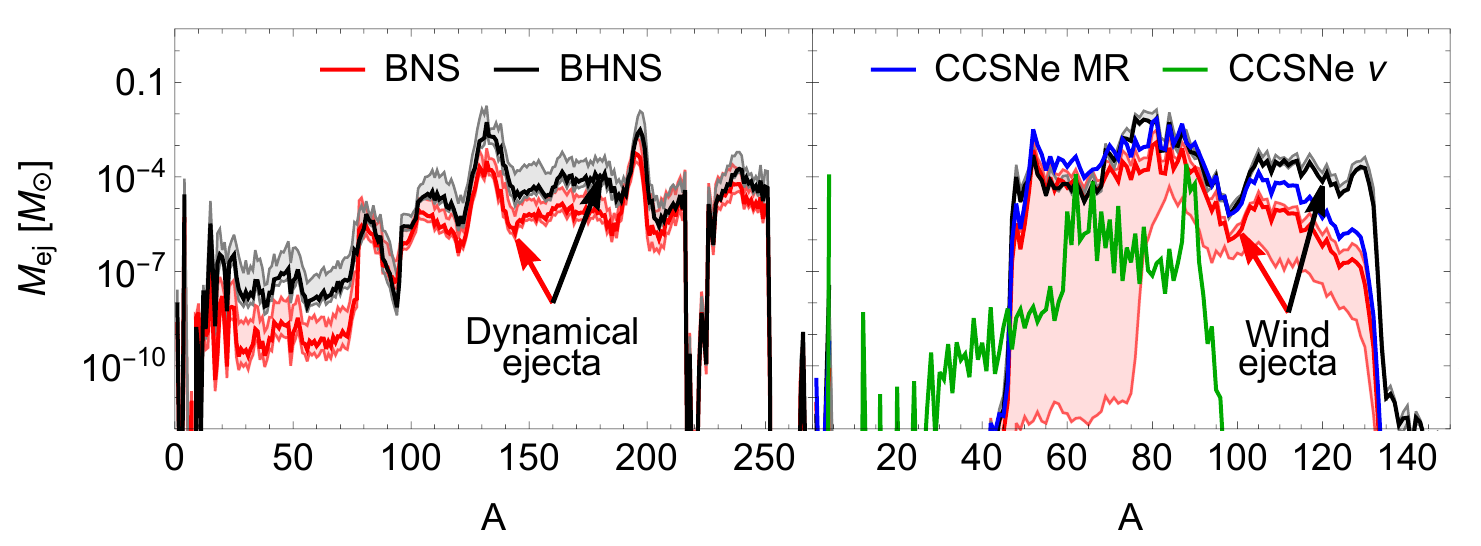}
\caption{We show the distributions of mass ejected as a function of nuclei mass number for all our typical scenarios. In both panels, we represent the BNS scenario with red, BHNS with black, MR CCSNe with blue, and thermal, neutrino-driven, CCSNe with green. Note that the mass number axes are different between the panels. \textit{Left-hand panel:} dynamical ejecta mass (in solar masses) per event for BNS and BHNS mergers. BHNS mergers may not produce dynamical ejecta if $R_{\rm ISCO}>R_{\rm TD}$. For the BNS case, we show the upper and lower limits in $M_{\rm ej}$ shaded in red, based on the $M_{\rm NS}$ parameter space: upper limit corresponds to ($M_{\rm NS,1}$, $M_{\rm NS,2}$) = ($1.2\,M_{\odot}$, $1.7\,M_{\odot}$) and lower limit is for $M_{\rm NS,1}=M_{\rm NS,2}=1.5\,M_{\odot}$. For the BHNS case, we also show the upper and lower limits shaded in gray, based on the $M_{\rm BH}-\chi_{\rm BH}$ parameter space. The upper limit is for a $12\,M_{\odot}$ BH with $\chi_{\rm BH}=0.97$ and the lower limit is for a $4\,M_{\odot}$ BH with $\chi_{\rm BH}=0.3$. \textit{Right-hand panel:} mass ejected in wind per event as a function of mass number for BNS and BHNS ejecta, MR CCSNe, and neutrino-driven CCSNe. Shaded areas reflect the limits based on the same parameter ranges as in the left-hand panel.} 
\label{fig:comparescenarios}
\end{figure*}

In the BNS and BHNS cases, the mass ejected is estimated using equations (\ref{bnsdynamical}) - (\ref{bhnsremnant}). To calculate the typical mass ejected in the BNS case, we assume that both NSs have mass $1.3\,M_{\odot}$ and radius $11.68\,{\rm km}$ (corresponding to $C_{\rm NS}=0.165$ for ALF2 EOS). If the BNS mass ratio is lower than 1 (in \citealt{Kruger:2020gig} the lower limit for fitting formula is $\sim0.775$), the dynamical and wind mass ejected could vary by a factor of $\sim10$ (see Appendix~\ref{massejectedappendix}). We assume a BNS mass ratio of 1 for simplicity. In the BHNS case, we assume again that the NS has a mass of $1.3\,M_{\odot}$ and radius $11.68\,{\rm km}$ (see \citealt{Kyutoku:2021icp} for EOS consistency with data and Appendix \ref{eosappendix} for how EOS affects detection prospects). The BH spin and mass also have an effect on the amount of mass ejected. We assume a Gaussian-like $M_{\rm BH}$ distribution as in \citet{Zhu:2020ffa} (their equation 6) which peaks around $M_{\rm BH} \approx 8\,M_{\odot}$. We assume that $\chi_{\rm BH}$ is distributed uniformly between 0 and 0.97. As the BH mass is not well constrained observationally, we adopt this mass distribution based on population synthesis studies. The BH spin is even less constrained, so the uniform distribution we assume is the simplest case to consider. A more detailed discussion on this is presented in $\S$\ref{discussion}. For the typical mass ejected in the BHNS scenario, we calculate the average mass ejected after performing a weighted integral over the $M_{\rm BH}$ and $\chi_{\rm BH}$ distributions (see also Appendix~\ref{massejectedappendix} for variation in BHNS mass ejected).

Lastly, CCSNe may eject more mass at smaller radii compared to the dynamical and wind ejecta from merger scenarios, so the initial densities calculated are too high for the {\tt SkyNet} EOS to work properly. To overcome this, we begin {\tt SkyNet} network calculations with an initial density of $10^9\,{\rm g~cm^{-3}}$. Assuming equation (\ref{densityevo}), it takes the expanding ejecta material less than $\sim1\,{\rm s}$ to attain this density value in the neutrino-driven and MR scenarios. Further, the temperature is still above the $\sim7\,{\rm GK}$ NSE threshold, so this approach should yield reasonable results.

\section{Nucleosynthesis results}\label{results}

In this section, we describe the results from nuclear reaction network calculations in three subsections: results from the merger scenarios, results from the core-collapse scenarios, and finally how these results can explain the observed stellar abundance patterns. 

\subsection{Merger scenarios}
In Fig.~\ref{fig:comparescenarios} we show the distribution of ejecta mass as a function of nuclei mass number from our nucleosynthesis analysis. In the left-hand panel, we show the distribution for the dynamical ejecta of our merger scenarios: BNS in red and BHNS in black. In the right-hand panel, we show the distribution of the wind ejecta for the merger scenarios along with the results from the CCSNe calculations (MR in blue and neutrino-driven in green). In shaded bands, we show how the distribution changes if we vary the initial BH and NS parameters. In both figure panels, the red-shaded bands show estimates for the upper (for $M_{\rm NS,1}=1.2\,M_{\odot}$ and $M_{\rm NS,2}=1.7\,M_{\odot}$) and lower (for $M_{\rm NS,1}=M_{\rm NS,2}=1.5\,M_{\odot}$) limits for the ejected mass. The difference is particularly striking for BNS wind ejecta; in this case, the total mass ejected is particularly low for two $1.5\,M_{\odot}$ NSs, but the maximum mass of nuclei synthesized is still similar. Also shown in both figure panels for BHNS (in grey shading) is an estimation of the upper (for $M_{\rm BH}=12\,M_{\odot}$ and $\chi_{\rm BH}=0.97$) and lower (for $M_{\rm BH}=4\,M_{\odot}$ and $\chi_{\rm BH}=0.3$) limits of the ejecta material. Note that for some BH mass and spin combinations there is no mass ejected, so the lower limit in these panels is approximately the least amount of mass ejected one could expect, given that there is some mass unbound from the system.

The dynamical ejecta material is neutron-rich ($Y_e=0.15/0.1$ for BNS/BHNS) enough to be able to undergo a robust r-process: less than the suggested threshold of $Y_e\sim0.25$ \citep{Kasen:2014toa,Lippuner:2015gwa}, while also having a low entropy and expansion time-scale. Although a low $S$ and $\tau_{\rm exp}$ generally leads to the synthesis of heavier nuclei, this relationship is not always monotonic, see fig.~4 of \citet{Lippuner:2015gwa}. In both BNS and BHNS mergers, the dynamical ejecta synthesizes up to and beyond the third r-process peak elements up to the actinides. The BNS case produces about as much first r-process peak nuclei as the BHNS case and underproduces intermediate nuclei, but otherwise the distributions share common features. The main difference between these two scenarios comes from the overall amount of mass ejected.

The BNS wind ejecta undergoes the `weak r-process' whereby nuclei are synthesized around the first r-process peak and up to $A\sim130$. Lower $Y_e$ values are required, however, to synthesize second r-process peak nuclei in a relative overabundance. In BHNS wind ejecta, where the winds are slightly more neutron rich ($Y_e\sim0.3$), the second-peak nuclei can be synthesized more readily.

Changing the initial BHNS parameters does not change the abundance pattern for dynamical ejecta much, as nuclei up to actinides are always synthesized. However, there are cases with high BH masses and low spins, where the BHNS merger does not successfully produce dynamical ejecta mass. It also affects the amount of mass ejected: between the ($4\,M_{\odot}$, $\chi_{\rm BH}=0.3$) and ($12\,M_{\odot}$, $\chi_{\rm BH}=0.97$) cases, there is a factor of $\sim10$ difference in total ejecta mass and a factor of $\sim10$ in abundance for the most massive nuclei synthesized in the dynamical ejecta.

\subsection{Core-collapse scenarios}

The MR CCSN and the BNS wind ejecta cases are quite similar in terms of r-process nuclei synthesized. As these are only moderately neutron-rich outflows (still more neutron-rich than the typical neutrino-driven core collapse), nuclei are synthesized in excess near the first r-process peak and beyond, up to mass numbers around $A\sim130$. MR CCSNe may be a relatively frequent and important site of weak r-process.

Since the thermal, neutrino-driven CCSN is only slightly neutron-rich, iron-like nuclei are synthesized but the distribution also peaks around nuclei with Z $\sim38-40$. In this slightly neutron-rich environment, subsequent alpha particle captures along the valley of stable nuclei - up to and past Fe - lead to an overabundance of nuclei with magic number N $=50$ (Sr, Y, and Zr). This produces the double-peak structure in green from the right-hand panel of Fig.~\ref{fig:comparescenarios} at iron and these N $=50$, or LEPP nuclei \citep{2011arcones}. LEPP elements can also be created under proton-rich conditions (see \citealt{2011arcones}) and also through the slow neutron-capture process (the s-process, see e.g. \citealt{Travaglio:2003qq}), all of which need to be considered for understanding GCE.

\subsection{Stellar abundance comparisons}

Stars like the Sun and metal-poor stars have been enriched over time by r-process events. The abundance data from these stars are useful to compare against our own nucleosynthesis yields. For solar abundances, we use the recent results from \citet{asplundabundances}, and for metal-poor stars, we include results from the recently measured HD222925 \citep{HD222925}. This particularly well-measured metal-poor star has abundance data for almost as many nuclei as the Sun and with similarly low error bars. The abundance distribution of this metal-poor star also matches well with the distributions of other metal-poor stars between the first and second r-process peak and around the third peak, with some exceptions \citep{Roederer:2022exr}.

\begin{figure}
\centering
\includegraphics[width=\linewidth]{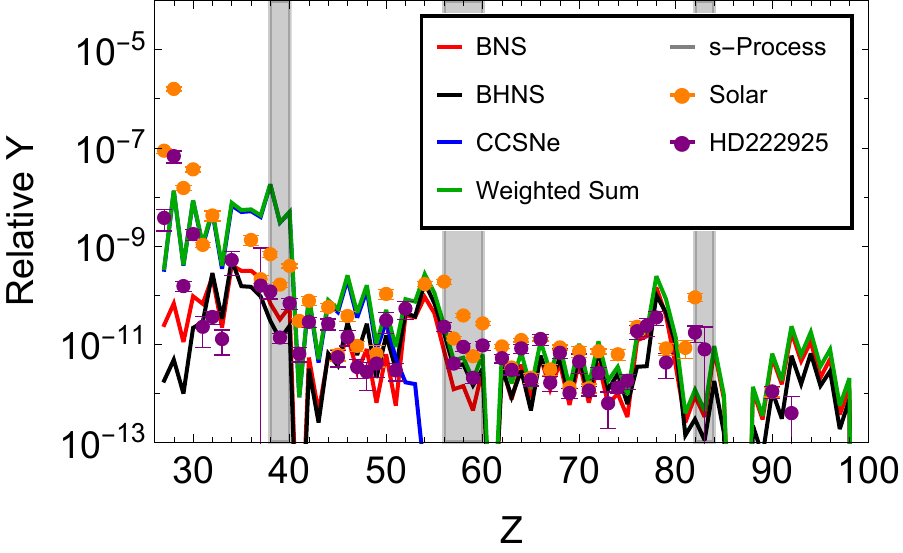}
\caption{Comparing solar abundance data (orange) from \citet{asplundabundances} and metal-poor star HD222925 (purple) \citep{HD222925} for trans-iron elements to our event-rate-weighted sum of nucleosynthesis yields (in green). Yields from our analysis and the HD222925 abundance data are rescaled to match the solar lanthanide (dysprosium, Z = 66) abundances. Generally, we find the agreement to be good. Intermediate elements are formed by stellar nucleosynthesis and s-process elements (at peaks of Z $\sim38-40$, $56-60$, $82-83$, shown in the grey bands), and these are not synthesized in our analysis.}
\label{fig:comparetodata}
\end{figure}

To weight the relative contribution of the sites considered here to the enhanced abundances found in the Sun and HD222925, we multiply the elemental abundances from Fig.~\ref{fig:comparescenarios} by the volumetric event rates of each scenario. Although the rates, especially of the merger scenarios, are very uncertain, the best rates come from GW analyses. The BNS volumetric event rates through GW analyses are ${\rm 320_{-240}^{+490}\,Gpc^{-3}~yr^{-1}}$ (${\rm 10-1700\,Gpc^{-3}~yr^{-1}}$) for GWTC-2 (GWTC-3) (\citealt{LIGOScientific:2020kqk,ligo2111.03634}). The BHNS volumetric event rates are ${\rm 7.8-140\,Gpc^{-3}~yr^{-1}}$ for GWTC-3 \citep{LIGOScientific:2020kqk}. From \citet{LIGObhnsrates}, the corresponding rate is ${\rm 45_{-33}^{+75}\,Gpc^{-3}~yr^{-1}}$ for two GW BHNS observations or ${\rm 130_{-69}^{+112}\,Gpc^{-3}~yr^{-1}}$ under broader assumptions for the BHNS masses. From \citet{Taylor:2014rlo}, the local CCSN rate is approximately ${\rm 1.06\times10^5\,Gpc^{-3}~yr^{-1}}$.

In Fig.~\ref{fig:comparetodata}, we show the result comparing the relative contributions to the elemental abundances for each scenario that we consider. The event-rate-weighted sum is shown in green assuming a BNS volumetric event rate of ${\rm 320\,Gpc^{-3}~yr^{-1}}$, a BHNS volumetric event rate of ${\rm 45\,Gpc^{-3}~yr^{-1}}$, and that MR CCSNe occur at around 1 per cent of the typical neutrino-driven CCSNe rate (\citealt{Nishimura:2015nca,Kashiyama:2015eua,Vlasov:2017nou,Halevi:2018vgp}). If MR CCSNe are associated with superluminous SNe, their rate is approximately 1 per cent of normal CCSNe \citep{Gomez:2022mdw}, whereas if MR CCSNe give rise to magnetar remnants, this number could be as high as $\gtrsim10$ per cent \citep{lewin2006,Beniamini:2019bga}. For a fair comparison, we rescale the relative abundances for the Sun (in orange) and HD222925 (in purple) to the lanthanide (specifically dysprosium, as in \citealt{Holmbeck:2022mog}) abundances of our event-rate-weighted sum. 

Overall, the solar and metal-poor abundances agree well across the distribution of elements with some exceptions (elements lighter than $\sim$iron and the s-process peak elements should not be considered in this comparison since we do not model synthesis of those nuclei). For one, the relative abundance of nuclei between $Z\sim30-35$ are underproduced in HD222925 compared to the Sun and our results. Interestingly, we see an overproduction of $Z\sim36-40$ nuclei coming primarily from both neutrino-driven and MR CCSNe. This may suggest that MR CCSNe occur less frequently than the estimated 1 per cent of typical CCSN rate. It could also mean that MR CCSNe are more neutron rich. However, the current rate and ejecta mass estimates would suggest that these systems produce significantly more r-process material than the typical merger scenarios if MR CCSNe were more neutron rich. There is also a slight overproduction of $Z\sim44-48$ elements from the MR CCSN and merger wind scenarios. The production of lanthanides is a particularly good match between all three abundance distributions. Finally, our analysis yields some overproduction of third r-process peak elements and a robust comparison to the actinide abundances is prevented by a lack of data for those nuclei. 

From this nucleosynthesis analysis, we see that BHNS and BNS mergers contribute almost equally to the production of robust r-process products (i.e., lanthanides, third-peak, and actinide production). The uncertainties in the rates of binary mergers and MR CCSNe are very large, so it is also useful to consider the prospects for observing mergers through joint GW and kilonova detections. Through multiple analysis channels, we can learn more about the merger scenarios and their relative contributions to r-process enrichment.

\section{Prospects for kilonova observability}\label{prospects}

\begin{figure*}
\includegraphics[width=\linewidth]{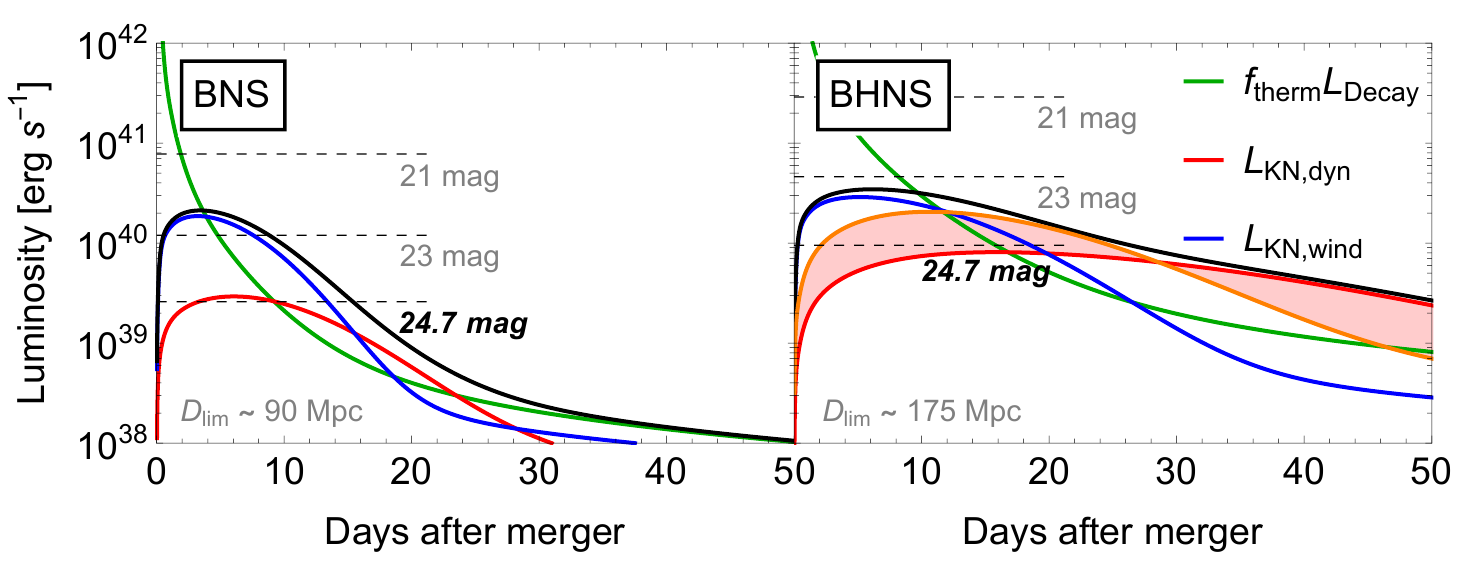}
\caption{Bolometric light curves for BNS and BHNS kilonovae, assuming a simple, one-zone emission model for the ejecta. In both panels, green curves are the luminosities from the decay of unstable nuclei multiplied by the thermalization fraction. In blue is the luminosity due to the lower opacity disc winds, in red is the luminosity due to the higher opacity dynamical ejecta, and in black is the sum of both these components. In the right-hand panel for BHNS case, the orange curve represents the luminosity if $\kappa \sim20~{\rm cm^2~g^{-1}}$ (as reported in \citealt{Tanaka:2019iqp}) instead of $\sim60~{\rm cm^2~g^{-1}}$ which we extrapolate from the exponential fit. Event rates are calculated by converting the bolometric luminosity into an $r$-band luminosity and assumes that the emission radiates as a blackbody. In both panels we also show, in dashed lines, the luminosity required to detect at least one event per year given the magnitude listed and limiting distance. Bolded is the estimated $r$-band magnitude for LSST and the listed $D_{\rm lim}$ is the distance at which these events would occur in order to have one event per year; see equation (\ref{dlim}) for details. In our fiducial BHNS kilonova light curve, the resulting luminosity is both brighter and longer than the corresponding BNS luminosity.}
\label{fig:knlightcurves}
\end{figure*}

Given the multimessenger detection of GW170817 including GRB 170817A and kilonova AT2017gfo (see e.g. \citealt{Pian:2017gtc,LIGOScientific:2017ync}), BNS mergers are confirmed as sites of r-process nucleosynthesis. Although it is more challenging to validate BHNS mergers as sites of robust r-process nucleosynthesis due to their non-detection of associated kilonova emission (see \citealt{Zhu:2020inc,Zhu:2020ffa,Chen:2021fro,Fragione:2021cvv}), the prospects are encouraging in the near future with the advent of advanced surveys. We generate simple kilonova estimates with a one-zone emission model by solving the following for $L_{\rm KN}$ (kilonova luminosity)
\begin{equation}
    \frac{dE(t)}{dt}=-L_{\rm KN}(t)-L_{\rm ad}(t)+f_{\rm therm}(t)L_{\rm decay}(t),
\end{equation}
where $E$ is the energy of the shell of ejecta mass $M_{\rm ej}$, $L_{\rm ad}=E(t)/\tau_{\rm dyn}$ is the adiabatic energy-loss rate ($\tau_{\rm dyn}=R_{\rm ej}/v_{\rm ej}$ is the dynamical time-scale, where $R_{\rm ej}$ is the radius of the ejecta and $v_{\rm ej}$ is the ejecta velocity, $R_{\rm ej}(t=0)=r_0$), $f_{\rm therm}$ is the thermalization efficiency of the r-process heating rate into the kilonova luminosity, and $L_{\rm decay}$ is the luminosity from the decay products of r-process nucleosynthesis. Here, $L_{\rm decay}=M_{\rm ej}\Dot{\epsilon}_{\rm decay}$ and the specific heating rate, $\Dot{\epsilon}_{\rm decay}$, is given directly from the {\tt SkyNet} calculations.

The estimated $L_{\rm KN}$ for each scenario requires knowledge of the mass ejected, velocity of the ejecta, and opacity, $\kappa$, of the heavy nuclei synthesized. The mass ejected is calculated as discussed earlier in \S\ref{methods}. For a binary mass ratio $Q$, the velocity of the BNS/BHNS dynamical ejecta can be estimated as \citep{Raaijmakers:2021slr}
\begin{equation}\label{velocityejecta}
    v_{\rm ej}=0.0149Q+0.1493.
\end{equation}
The BHNS wind ejecta velocity is estimated to be $\sim0.15\,c$ ($Q=6.2$ for the average BH mass in our distribution and a $1.3\,M_{\odot}$ NS), and the BNS wind ejecta velocity is estimated to be $\sim0.07\,c$ (average of the neutrino-driven wind velocity, $\sim0.08\,c$, and viscous-driven wind velocity, $\sim0.06\,c$, see \citealt{Radice:2018pdn}).

For $\kappa$, we use the results from \citet{Tanaka:2019iqp} for $Y_e\sim0.15-0.4$ ($\kappa$ for $Y_e=0.1$ is likely underestimated due to a lack of atomic data for the actinides). For the typical BHNS dynamical ejecta and the thermal CCSNe cases, we estimate $\kappa$ in this $Y_e$ range by fitting an exponential function to the available opacity data to get $\kappa=211.98\exp(-12.33~Y_e)\, {\rm cm^2~g^{-1}}$.

In general, the thermalization fraction, $f_{\rm therm}$, is a time-dependent quantity that describes the fraction of energy that is absorbed by the ejecta from radioactive processes that includes fission fragments, $\alpha$ particles, $\beta$ particles, and $\gamma$-rays. We assume that the total thermalization efficiency can be approximately described by \citep{Barnes:2016umi}
\begin{equation}\label{thermfrac}
    f_{\rm therm}=0.36\left[\exp(-xt)+\frac{\ln(1+2yt^z)}{2yt^z}\right],
\end{equation}
where $x,~y,$ and $z$ are fit coefficients. The fit coefficients are given in Table 1 of \citet{Barnes:2016umi}; we choose the fit coefficients based on the ejecta mass/velocity for the `Random' column, that agree best with their $M_{\rm ej}/v_{\rm ej}$ models for each merger scenario (see also \citealt{Rosswog:2016dhy}).

In Fig.~\ref{fig:knlightcurves}, we show the kilonova light curves for the BNS (left-hand) and BHNS (right-hand) merger scenarios. In green, we show the luminosity of the radioactive decay products. In blue, we show the luminosity from the wind ejecta, in red we show the luminosity from the dynamical ejecta, and in black we show the sum of blue and red components. In the left-hand panel, we see that the results are qualitatively consistent with other BNS studies. In the first $\sim5\,{\rm d}$ after merger, there is a peak corresponding to higher $Y_e$ (lower $\kappa$) wind ejecta followed by a lower and broader peak corresponding to lower $Y_e$ (higher $\kappa$) dynamical ejecta. This is expected as the early blue kilonova component is followed by the red component at late times for AT2017gfo (see e.g. \citealt{Pian:2017gtc}). 

The bolometric luminosity for AT2017gfo is generally systematically higher than our simple one-zone model estimate (see e.g. \citealt{Smartt:2017fuw,Waxman:2017sqv,Arcavi:2018mzm,Tanaka:2019iqp} for some comparisons). This is largely due to the inferred opacities, but there are also notable differences in inferred $Y_e$, $v_{\rm ej}$, heating rates, and $M_{\rm ej}$ (where $M_{\rm ej}$ is a factor of $\sim10$ higher). Further, AT2017gfo may be consistent with an additional luminosity source from the cooling of ejecta that has been heated either by winds, the jet or the cocoon \citep{Arcavi:2018mzm}. Lastly, this may also be explained by the simplicity of our model, although \citet{Tanaka:2017lxb} note that their simpler model is similar to their more realistic models if the thermalization efficiencies are accounted for. In our BNS model, the red component of the kilonova does not contribute significantly to the luminosity until around $15\,{\rm d}$ after the merger.

Successful BHNS mergers, as shown in the right-hand panel of Fig.~\ref{fig:knlightcurves}, also show a higher $Y_e$ (lower $\kappa$) component at early times, followed by a lower $Y_e$ (higher $\kappa$) component. However, the time-scales here are longer as compared to the BNS scenario. The red component does not contribute significantly to the kilonova luminosity until around $20\,{\rm d}$ after the merger, but it is brighter and longer than the corresponding component for BNS mergers. We also plot, in orange, $L_{\rm KN,dyn}$ for $\kappa\sim20\,{\rm cm^2~g^{-1}}$ (as reported in \citealt{Tanaka:2019iqp}) for $Y_e=0.1$, however this value may not be accurate due to the incomplete atomic data for actinides (the lower $\kappa$ case would lead to an even brighter, but shorter, kilonova). This results in an altogether brighter and longer kilonova light curve for a typical, successful, BHNS merger. While the BNS light curve drops by an order of magnitude after roughly $20\,{\rm d}$, the BHNS light curve drops by an order of magnitude after 50 or more days. In part, this is due to the increased opacity for very neutron rich outflows, but primarily comes from the prolonged heating rate in BHNS mergers. This makes BHNS kilonovae an intriguing source with different cadence requirements compared to BNS kilonovae. Whereas BNS kilonovae require a cadence of $\lesssim1-2\,{\rm d}$ (\citealt{Zhu:2021zmy,Zhu:2021ram}), BHNS kilonovae should require a cadence of $\lesssim1-3\,{\rm d}$ for multiple observing epochs \citep{Zhu:2020ffa}.

We estimate the detection rates of these kilonova transients. In particular, we are interested in the number of kilonova events detectable by the LSST of the Vera C. Rubin Observatory (see \citealt{LSST:2008ijt,LSSTScience:2009jmu}), which is proposed to have system first light in July 2024. The prospect of GW detection during the LIGO/Virgo/KAGRA O4 observation run (proposed to begin in May 2023 and run for 18 months) is also useful. In LIGO O4 run, the limiting distance, $D_{\rm lim,GW}$ for the detection of BNS/BHNS mergers is predicted to be $160-190/300-330\,{\rm Mpc}$ \citep{KAGRA:2013rdx}. The magnitude-limited distance for Rubin observatory is estimated from
\begin{equation}\label{dlim}
    D_{\rm lim,KN}=\left(\frac{L_{\rm \nu,peak}}{4\pi F_{\rm lim}}\right)^{1/2},
\end{equation}
where $L_{\rm \nu,peak}$ is the maximum kilonova luminosity (in ${\rm erg~s^{-1}~Hz^{-1}}$) at frequency $\nu$ of the observing band, and $F_{\rm lim}=(3631~{\rm Jy}\times 10^{-m_{\rm lim}/2.5})$ is the limiting flux density corresponding to a limiting magnitude $m_{\rm lim}$. Assuming that the luminosity radiates as a blackbody, $L_{\nu}$ is given by
\begin{equation}
    L_{\nu}=\frac{2\pi h\nu^3}{c^2}\frac{1}{\exp(h\nu/k_BT_{\rm pho})-1}R_{\rm pho}^2d\Omega,
\end{equation}
where $T_{\rm pho}$ and $R_{\rm pho}$ are the temperature and radius of the photosphere, respectively. This blackbody assumption is reasonable, especially within the first $\sim5-10\,{\rm d}$, where the luminosity peaks (\citealt{Smartt:2017fuw,Waxman:2017sqv,Arcavi:2018mzm}). We follow \citet{Villar:2017wcc} to model $T_{\rm pho}$ and $R_{\rm pho}$ (their equation 4 and 5) and use the best-fitting values of the temperature floor, $T_c$, from their two-component model. We use the blue $T_c$ value for our wind ejecta and red $T_c$ value for our dynamical ejecta. Since these are best-fitting values based on GW170817, these temperatures are reasonable for BNS mergers but we also assume them for BHNS mergers. This temperature floor occurs near the recombination temperature of the lanthanides \citep{Barnes:2013wka}; since both BNS and BHNS mergers similarly undergo a robust r-process (including the synthesis of lanthanides) this assumption may be reasonable.

Assuming an $r$-band ($\nu\sim600\,{\rm nm}$) $5\sigma$ limiting magnitude of 24.7 with $30\,{\rm s}$ exposure time \citep{LSSTScience:2009jmu}, $D_{\rm lim,KN}\sim 170\,{\rm Mpc}$ for BNS mergers and $\sim 210\,{\rm Mpc}$ for BHNS mergers. Since $D_{\rm lim,KN}<D_{\rm lim,GW}$, GWs can be used as a trigger preceding kilonova detection. We then estimate the kilonova detection rate as
\begin{equation}
    \Dot{N}_{\rm KN}=\frac{4\pi}{3}D_{\rm lim,KN}^3\mathcal{R},
\end{equation}
where $\mathcal{R}$ is the volumetric event rate for BNS/BHNS mergers. Adopting the most probable volumetric event rate values with the maximal luminosities from Fig.~\ref{fig:knlightcurves}, we estimate that $\sim7$ BNS mergers and $\sim2$ BHNS mergers will be detectable per year by LSST - almost all of which can, in principle, be triggered by GW detections if there is an observational overlap. In Fig.~\ref{fig:knlightcurves} we also plot several dashed curves for both BNS and BHNS mergers. These lines represent the minimum bolometric luminosity required for at least one detection per year, assuming the magnitude listed (using equation \ref{dlim}). A magnitude of 24.7 is bolded to represent the r-band limiting magnitude for LSST and these calculations assume that $D_{\rm lim,KN}\sim90\,{\rm Mpc}$ for BNS mergers and $\sim175\,{\rm Mpc}$ for BHNS mergers. The BNS and BHNS kilonova event rates are similar to those predicted in \citet{Zhu:2021ram} and \citet{Gupta:2023evt}, respectively, where similar volumetric rate, EOS, and LSST observation assumptions are made (and both include more in-depth studies on the prospects for GW detection as a trigger for target-of-opportunity multimessenger searches).

\section{Discussion}\label{discussion}

The abundance distribution of nuclei synthesized in the dynamical and wind ejecta of BHNS and BNS mergers are very similar. Given their typical mass ejected ($\sim 0.7\times 10^{-2} M_\odot$ for BNS and $\sim 9\times 10^{-2} M_\odot$ for BHNS) and the most likely values of the local volumetric event rates used in this study (${\rm 320\,Gpc^{-3}~yr^{-1}}$ for BNS and ${\rm 45\,Gpc^{-3}~yr^{-1}}$ for BHNS), this suggests that these two merger scenarios contribute similar amounts to the observed abundance of third r-process peak nuclei and heavier nuclei. However, as these rate measurements are uncertain by around 100 per cent (or more under broader assumptions, see \citealt{LIGOScientific:2021qlt}), better measurements could lead to different conclusions about which merger scenario dominantly enriches stars with robust r-process nuclei. 

Further, this analysis assumes that the relative rates between BNS and BHNS mergers are similar to what they are locally. Some studies suggest that these merger rates depend on redshift and metallicity such that in low-metallicity environments, the BHNS merger rate is actually higher than the BNS merger rate. This would alter the evolution history of enrichment of stars with heavy nuclei, and also, in principle, change the distribution of BH masses and spins (see \citealt{Kobayashi:2022qlk}, and sources therein). To check the robustness of the local rate assumption, we look at the redshift evolution of BHNS mergers within a redshift of $\sim0.05$ (corresponding to the proposed LIGO O4 BNS sensitivity range and LSST's magnitude-limited distance of $\sim200\,{\rm Mpc}$) for the three models in \citet{Zhu:2020ffa}. The log-normal delay model results in, at most, a $\sim20$ per cent increase in the rate up to a redshift of 0.05, so there could be a modest increase in the event rate of BHNS mergers. Additionally, knowing the delay-time distributions for compact object mergers is integral for understanding relative contributions from different r-process sites but also if mergers can explain the heavy element enrichment early on in the evolution of some galaxies. This tension is recently discussed in, e.g., \citet{Simon:2022prp} but population synthesis models (e.g., \citealt{Dominik:2012kk,Mennekens:2016jcs,Santoliquido:2022kyu}) and some observational constraints (\citealt{Kim:2006fm,Beniamini:2019iop,Simon:2022prp}) do allow for mergers to happen on time-scales of the order $\sim100\,{\rm Myr}$ or less. Because these rates are uncertain, model dependent, and short delay time distributions are possible, we assume volumetric event rates found by LIGO studies to compare against stellar abundance distributions in Fig.~\ref{fig:comparetodata} and calculate detection rates.

Whether there are heavy elements ejected and synthesized in BHNS mergers also depends on whether or not there is a successful outflow, so understanding the BH mass and spin distribution is important. For the BH mass distribution, we choose the functional form from \citet{Zhu:2020ffa}, which is based on the population synthesis models of \citet{Giacobbo:2018etu}. The inferred BH masses from two observed BHNS mergers fall within the range of this assumed distribution and seem to be consistent with current population synthesis models \citep{LIGOScientific:2021qlt}. For the BH spin, we assume that it is distributed uniformly between 0 and 0.97. The distribution of BH spins could be quasi-bimodal: low-spin values if BHNS systems evolve similar to BBH systems \citep{LIGOScientific:2018mvr}, or high-spin values if some X-ray binaries evolve into BHNS systems (see \citealt{Zhu:2020ffa}, and sources therein), but this could be subject to observational biases. Further, the BHNS mergers observed may be consistent with intermediate spin values \citep{LIGOScientific:2021qlt}, so for simplicity, we assume the uniform distribution. For the assumed BH mass and spin distributions in this work, the fraction of BHNS systems that successfully unbind mass is about $\sim80$ per cent. If we assume the bimodal BH spin distribution from \citet{Zhu:2020ffa}, that fraction decreases to about $\sim61$ per cent. Correspondingly, the number of detectable BHNS kilonovae may decrease by around 20 per cent if the spins are naturally bimodal. Our fractions for BHNS systems that eject mass successfully (including in Appendix~\ref{eosappendix}) roughly agree with the more optimistic estimates in the literature, which puts this number anywhere between $\lesssim1$ and $70$ per cent (see \citealt{Roman-Garza:2020uou,Broekgaarden:2021iew,Fragione:2021cvv,Drozda:2020qab,Biscoveanu:2022iue}).

Successful observation of kilonovae also depends on the geometry of the ejecta \citep{Kawaguchi:2016ana} and the viewing angle. Since the dynamical ejecta is anisotropic, the relative angle at which the kilonova is observed results in some variation in the luminosity (a factor of $\sim2-3$ difference in \citealt{Zhu:2020inc} for BNS mergers). However, the brightest component of the kilonova comes from the wind ejecta in our models which is more isotropic, so the viewing angle effect is relatively minor in terms of luminosity. Additionally, a recent analysis of GW170817/AT2017gfo proposed that the kilonova emission was likely to be highly spherical \citep{Sneppen:2023vkk}. Viewing angle also has implications for the ability to distinguish between kilonovae and off-axis GRB afterglows, which may peak in luminosity at similar times. \citet{Zhu:2021zmy} found that the light curves are dominated by kilonovae only at large viewing angles for BNS mergers, so more robust techniques, like observing in multiple wavelengths, will be required for distinguishing between kilonovae and GRB afterglows. The dynamical ejecta of compact binary mergers may also give rise to a kilonova afterglow (see the analytical models of \citealt{Bhattacharya:2018lmw,Kathirgamaraju:2019xwu,Sadeh:2022enp,nedora2023} and the numerical models of \citealt{combi2022,Nedora:2023hiz}). There is increasing observational evidence of a kilonova afterglow associated with GW170817/AT2017gfo (see \citealt{Nedora:2021eoj,Hajela:2021faz}), which peaks $\sim 3.5\,{\rm yr}$ after the initial merger.

We assume the NS EOS to be ALF2, which is consistent with current observations, but there are additional EOS (both stiffer and softer, see e.g. fig. 11 of \citealt{Kyutoku:2021icp}), which can play a role in determining the amount of mass ejected (see e.g. \citealt{Radice:2018pdn}, for a quantitative discussion of this in the context of BNS mergers), and also, whether or not BHNS mergers successfully unbind mass from the NS. Although this would certainly play a role in the dynamics of CCSNe as well, our parametrized study does not directly depend on the EOS for supernovae. In Appendix~\ref{eosappendix}, we show the resulting light curves from varying the EOS and its effect on the event rates. The bolometric light curves around the peak luminosity are similar (with the exception of the APR4 EOS curve in BNS mergers), so the resulting variation in event rates is small and the uncertainty in event rates is dominated by the volumetric occurrence rates.

The initial thermodynamic and outflow conditions are chosen as `typical' values out of a range of parameter space that is primarily motivated by  simulations. $Y_e$ plays the largest role in determining the final abundance distribution. However, the distribution of electron fractions for each scenario is somewhat unclear. The $Y_e$ distributions for the BHNS wind is broad, which could imply BHNS outflows do not synthesize as much abundance in lighter nuclei compared to their BNS counterparts. The distribution for MR CCSNe is also particularly broad and might imply that MR CCSNe should also contribute to the production of robust r-process nuclei ($A\sim200-240$). Additionally, the electron fraction varies in different parts of all of these outflows, as seen in GRMHD simulations; we take the values of $Y_e$ that are reasonable and likely to give us some r-process nuclei but there will be lighter mass contributions as well. These systems should also have a distribution in ejected velocity which can affect the shape of the kilonova light curve. For instance, if the mass elements follow a distribution that is skewed toward slower velocities as in fig. 32 of \citet{Radice:2018pdn}, we expect the kilonova light curve to peak later and evolve slower. Our two-component (dynamical and wind ejecta) model in part represents the variation in electron fraction and ejecta velocities for the purposes of this study. Finally, the abundance distributions of nuclei synthesized, shown in Fig.~\ref{fig:comparescenarios}, generally agree with the abundance distributions found in simulations (see the references in Table~\ref{table:typicalparameterrange}), even in the MR CCSNe case where more complex density profiles may arise.

There are a number of secondary uncertainties that can contribute to the detectability of mergers as well. This includes the fit equations to numerical relativistic simulations. The  equations for ejecta mass that we use provide reasonable fits to the simulation results, especially when compared to other fitting formulae \citep{Henkel:2022naw}, provided the BH and NS initial conditions used are within the range of acceptable parameters. Uncertainties in nuclear processes, such as beta decay rates, can have a notable impact on the heating rate, and thus, the brightness of the kilonova (\citealt{Kullmann:2022xuj,Lund:2022bsr}). Finally, the opacity of r-process ejecta is fairly uncertain. For example, the $\kappa$ assumed in \citet{Waxman:2017sqv} is much lower than what is assumed in this work and supported by atomic structure and radiative transfer simulations as in \citet{Tanaka:2019iqp} (see also the opacities used in \citealt{Raaijmakers:2021slr,Lund:2022bsr}). Here, we choose to adopt the models of \citet{Tanaka:2019iqp}, which are averaged across r-process nuclei and determined in the context of kilonova emission. These are valid at high temperatures and early times when the peak kilonova luminosity is achieved, and are therefore reasonable when calculating the rate of detectable events.

\begin{figure}
\centering
\includegraphics[width=\linewidth]{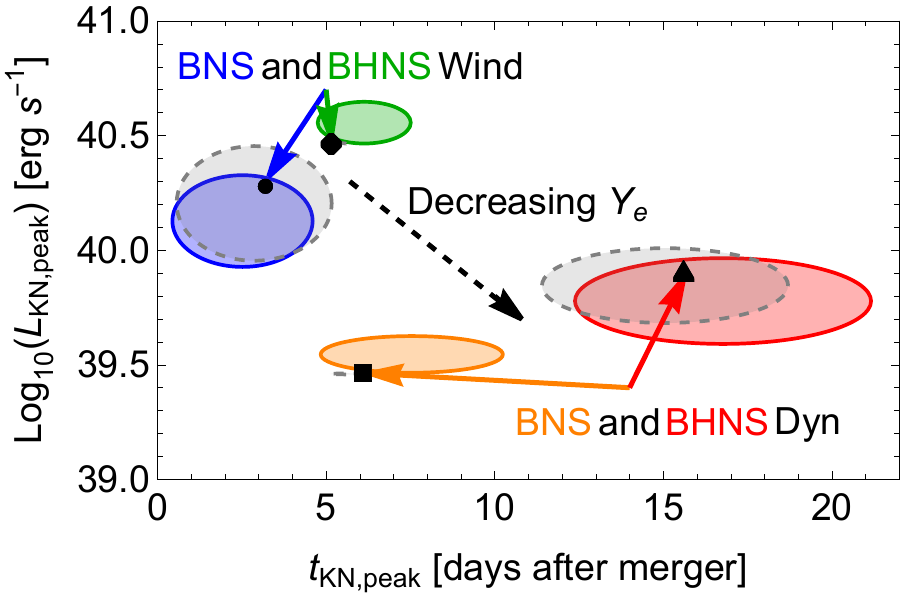}
\caption{The data points represent the kilonova peak luminosity and time of peak luminosity ($L_{\rm KN,peak}/t_{\rm KN,peak}$) values from our fiducial model parameters. Blue/green represents BNS/BHNS wind ejecta and orange/red represents BNS/BHNS dynamical ejecta. The correspondingly coloured bubbles show the range of values when varying initial BH and NS parameters. For BHNS, our upper wind/dynamical ejecta limit is given by ($M_{\rm BH}$, $\chi_{\rm BH}$) = ($4\,M_{\odot}$, 0.97)/($12\,M_{\odot}$, 0.97) and the lower limit for wind and dynamical ejecta is given by ($M_{\rm BH}$, $\chi_{\rm BH}$) = ($8\,M_{\odot}$, 0.4). For BNS mergers, we find the upper limit by evaluating the mass ejected for $M_{\rm NS,1}=1.2\,M_{\odot}$ and $M_{\rm NS,2}=1.7\,M_{\odot}$ and the lower limit by evaluating the mass ejected for $M_{\rm NS,1/2}=1.5\,M_{\odot}$. In dashed grey we show the ranges in $L_{\rm KN,peak}$ and $t_{\rm KN,peak}$ by varying the EOS (see Appendix~\ref{eosappendix}). The `Decreasing $Y_e$' arrow qualitatively shows how the luminosities change as a function of $Y_e$: with decreasing $Y_e$, opacities increase leading to a dimmer and more delayed kilonova.}
\label{fig:lpeaktpeak}
\end{figure}

Our estimated kilonova rates detectable with LSST are dependent directly on the peak luminosity of each merger scenario. In Fig.~\ref{fig:lpeaktpeak} we show our fiducial values of the peak bolometric luminosity, $L_{\rm peak}$, and time when peak luminosity is attained, $t_{\rm peak}$ as data points. The bubbles approximately correspond to the uncertainty in each scenario, found by varying the initial BH and NS mass, the BH spin parameters, and NS EOS from Fig.~\ref{fig:schematic} and Table~\ref{table:typicalparameterrange}. In blue/green we show the peak luminosity ($L_{\rm peak}$) and time of peak luminosity ($t_{\rm peak}$) range for the BNS/BHNS wind ejecta and in orange/red we show the range for BNS/BHNS dynamical ejecta. For BNS mergers, we find the upper limit by evaluating the mass ejected for $M_{\rm NS,1}=1.2\,M_{\odot}$ and $M_{\rm NS,2}=1.7\,M_{\odot}$ and the lower limit by evaluating the mass ejected for $M_{\rm NS,1/2}=1.5\,M_{\odot}$. For BHNS mergers, our upper limit for wind ejecta is given by a $4\,M_{\odot}$ BH with $\chi_{\rm BH}=0.97$ and the upper limit for dynamical ejecta is given by a $12\,M_{\odot}$ BH with $\chi_{\rm BH}=0.97$. The lower limit is for a $8\,M_{\odot}$ BH with $\chi_{\rm BH}=0.4$. It can be seen that the bubbles do not overlap, and therefore, the BHNS mergers tend to be consistently brighter and peak later than BNS mergers. In the dashed, grey bubbles, we show the range in $L_{\rm peak}$ and $t_{\rm peak}$ after varying the NS EOS (see Appendix~\ref{eosappendix}). The EOS variation is in general lesser than the colored bubbles. This is especially true for the BHNS wind and BNS dynamical ejecta. If our fiducial $Y_e$ values were different for each of our model components, we would see shifts of the bubbles as well, i.e., if our wind components were more neutron rich, they would synthesize heavier nuclei, give rise to higher opacities, and produce a dimmer and more delayed signal. This variation in brightness and timing, shown qualitatively with the dashed arrow in Fig.~\ref{fig:lpeaktpeak}, is substantial ($\sim10^{40}-10^{42}\,{\rm erg~s^{-1}}$ for $L_{\rm peak}$ and $\sim5-20\,{\rm d}$ for $t_{\rm peak}$), so estimating $Y_e$ more accurately is especially important for distinguishing different ejecta components.

\section{Summary}\label{summary}

We performed a comparative nucleosynthesis analysis for the outflows of BNS and BHNS mergers as well as MR and thermal CCSNe. We conclude that nuclei abundance distributions are overall similar between these two merger scenarios. Based on the current, but poorly constrained event rates, BNS and BHNS mergers may both contribute significantly to the amount of r-process enrichment in stars including our Sun and metal-poor stars. When comparing our nucleosynthesis yields to stellar abundances, there is an overproduction of LEPP nuclei ($Z\sim36-40$) with the inclusion of MR CCSNe. This could suggest that MR CCSNe tend to occur less frequently than 1 per cent of the typical CCSN rate or that they are even more neutron-rich than what is considered here. Compared to compact object mergers like BNS and BHNS that synthesize mass numbers up to $A\sim200-240$, MR CCSNe typically only synthesize up to $A\sim140$ and thermal CCSNe typically only synthesize up to $A\sim80-90$. MR CCSNe could synthesize higher mass numbers, though, if their outflows are more neutron-rich than considered here.

We also estimated the kilonova light curves resulting from the decay of unstable r-process nuclei for the merger scenarios and finally conclude that the prospects for determining which merger scenario contributes more to stellar r-process enrichment are encouraging for the next few years. This is especially true if both LIGO (O4 observation run) and LSST are collecting data at the same time. The LSST magnitude-limited-distance is also similar to the projected distance sensitivity for LIGO. In the $r$-band, LSST should be able to see $\sim7$ BNS and $\sim2$ BHNS events per year, which will be extremely important to better understand a variety of open astrophysics questions. Further, observatories like LSST could observe the first BHNS kilonova event, which is characterized by a brighter and longer light curve compared to its BNS counterpart.

\section*{Acknowledgements}
We thank David Radice and Masaomi Tanaka for useful discussions. We also thank Masaomi Tanaka for providing the code to generate kilonova light curves. We thank Chris Ashall and Ashley Villar for carefully reading the manuscript and providing insightful comments which have helped to improve the paper. 
NE\ and MB\ are supported by NSF grant no.\ AST-1908960. MB\ acknowledges support from Eberly Research Fellowship at the Pennsylvania State University. The work of SH\ is supported by the U.S.\ Department of Energy Office of Science under award number DE-SC0020262, NSF Grant No.\ AST-1908960, No.\ PHY-1914409 and No.~PHY-2209420, and JSPS KAKENHI Grant No. JP22K03630. This work was supported by World Premier International Research Center Initiative (WPI Initiative), MEXT, Japan.

\section*{Data availability}
The data underlying this article will be shared on reasonable request to the corresponding author.

\bibliography{main}

\begin{thebibliography}{}
\makeatletter
\relax
\def\mn@urlcharsother{\let\do\@makeother \do\$\do\&\do\#\do\^\do\_\do\%\do\~}
\def\mn@doi{\begingroup\mn@urlcharsother \@ifnextchar [ {\mn@doi@}
  {\mn@doi@[]}}
\def\mn@doi@[#1]#2{\def\@tempa{#1}\ifx\@tempa\@empty \href
  {http://dx.doi.org/#2} {doi:#2}\else \href {http://dx.doi.org/#2} {#1}\fi
  \endgroup}
\def\mn@eprint#1#2{\mn@eprint@#1:#2::\@nil}
\def\mn@eprint@arXiv#1{\href {http://arxiv.org/abs/#1} {{\tt arXiv:#1}}}
\def\mn@eprint@dblp#1{\href {http://dblp.uni-trier.de/rec/bibtex/#1.xml}
  {dblp:#1}}
\def\mn@eprint@#1:#2:#3:#4\@nil{\def\@tempa {#1}\def\@tempb {#2}\def\@tempc
  {#3}\ifx \@tempc \@empty \let \@tempc \@tempb \let \@tempb \@tempa \fi \ifx
  \@tempb \@empty \def\@tempb {arXiv}\fi \@ifundefined
  {mn@eprint@\@tempb}{\@tempb:\@tempc}{\expandafter \expandafter \csname
  mn@eprint@\@tempb\endcsname \expandafter{\@tempc}}}

\bibitem[\protect\citeauthoryear{Abbott et~al.}{Abbott
  et~al.}{2017a}]{LIGOScientific:2017vwq}
Abbott B.~P.,  et~al., 2017a, \mn@doi [Phys. Rev. Lett.]
  {10.1103/PhysRevLett.119.161101}, 119, 161101

\bibitem[\protect\citeauthoryear{Abbott et~al.}{Abbott
  et~al.}{2017b}]{LIGOScientific:2017ync}
Abbott B.~P.,  et~al., 2017b, \mn@doi [Astrophys. J. Lett.]
  {10.3847/2041-8213/aa91c9}, 848, L12

\bibitem[\protect\citeauthoryear{Abbott et~al.}{Abbott
  et~al.}{2017c}]{LIGOScientific:2017zic}
Abbott B.~P.,  et~al., 2017c, \mn@doi [Astrophys. J. Lett.]
  {10.3847/2041-8213/aa920c}, 848, L13

\bibitem[\protect\citeauthoryear{Abbott et~al.}{Abbott
  et~al.}{2018}]{KAGRA:2013rdx}
Abbott B.~P.,  et~al., 2018, \mn@doi [Living Rev. Rel.]
  {10.1007/s41114-020-00026-9}, 21, 3

\bibitem[\protect\citeauthoryear{Abbott et~al.}{Abbott
  et~al.}{2019}]{LIGOScientific:2018mvr}
Abbott B.~P.,  et~al., 2019, \mn@doi [Phys. Rev. X]
  {10.1103/PhysRevX.9.031040}, 9, 031040

\bibitem[\protect\citeauthoryear{Abbott et~al.}{Abbott
  et~al.}{2021a}]{LIGOScientific:2020kqk}
Abbott R.,  et~al., 2021a, \mn@doi [Astrophys. J. Lett.]
  {10.3847/2041-8213/abe949}, 913, L7

\bibitem[\protect\citeauthoryear{Abbott et~al.}{Abbott
  et~al.}{2021b}]{LIGObhnsrates}
Abbott R.,  et~al., 2021b, \mn@doi [Astrophys. J. Lett.]
  {10.3847/2041-8213/ac082e}, 915, L5

\bibitem[\protect\citeauthoryear{Abbott et~al.}{Abbott
  et~al.}{2021c}]{LIGOScientific:2021qlt}
Abbott R.,  et~al., 2021c, \mn@doi [Astrophys. J. Lett.]
  {10.3847/2041-8213/ac082e}, 915, L5

\bibitem[\protect\citeauthoryear{Akmal \& Pandharipande}{Akmal \&
  Pandharipande}{1997}]{PhysRevC.56.2261}
Akmal A.,  Pandharipande V.~R.,  1997, \mn@doi [Phys. Rev. C]
  {10.1103/PhysRevC.56.2261}, 56, 2261

\bibitem[\protect\citeauthoryear{Alford, Braby, Paris  \& Reddy}{Alford
  et~al.}{2005}]{ALF2EOS}
Alford M.,  Braby M.,  Paris M.~W.,   Reddy S.,  2005, \mn@doi [Astrophys. J.]
  {10.1086/430902}, 629, 969

\bibitem[\protect\citeauthoryear{{Aoki}, {Ishimaru}, {Aoki}  \&
  {Wanajo}}{{Aoki} et~al.}{2017}]{2017ApJ...837....8A}
{Aoki} M.,  {Ishimaru} Y.,  {Aoki} W.,   {Wanajo} S.,  2017, \mn@doi [\apj]
  {10.3847/1538-4357/aa5d08}, \href
  {https://ui.adsabs.harvard.edu/abs/2017ApJ...837....8A} {837, 8}

\bibitem[\protect\citeauthoryear{Arcavi}{Arcavi}{2018}]{Arcavi:2018mzm}
Arcavi I.,  2018, \mn@doi [Astrophys. J. Lett.] {10.3847/2041-8213/aab267},
  855, L23

\bibitem[\protect\citeauthoryear{{Arcones} \& {Montes}}{{Arcones} \&
  {Montes}}{2011}]{2011arcones}
{Arcones} A.,  {Montes} F.,  2011, \mn@doi [\apj] {10.1088/0004-637X/731/1/5},
  \href {https://ui.adsabs.harvard.edu/abs/2011ApJ...731....5A} {731, 5}

\bibitem[\protect\citeauthoryear{Arcones \& Thielemann}{Arcones \&
  Thielemann}{2023}]{Arcones:2022jer}
Arcones A.,  Thielemann F.-K.,  2023, \mn@doi [Astron. Astrophys. Rev.]
  {10.1007/s00159-022-00146-x}, 31, 1

\bibitem[\protect\citeauthoryear{Arnould, Goriely  \& Takahashi}{Arnould
  et~al.}{2007}]{Arnould:2007gh}
Arnould M.,  Goriely S.,   Takahashi K.,  2007, \mn@doi [Phys. Rept.]
  {10.1016/j.physrep.2007.06.002}, 450, 97

\bibitem[\protect\citeauthoryear{{Asplund}, {Amarsi}  \& {Grevesse}}{{Asplund}
  et~al.}{2021}]{asplundabundances}
{Asplund} M.,  {Amarsi} A.~M.,   {Grevesse} N.,  2021, \mn@doi [\aap]
  {10.1051/0004-6361/202140445}, \href
  {https://ui.adsabs.harvard.edu/abs/2021A&A...653A.141A} {653, A141}

\bibitem[\protect\citeauthoryear{Bardeen, Press  \& Teukolsky}{Bardeen
  et~al.}{1972}]{Bardeen:1972fi}
Bardeen J.~M.,  Press W.~H.,   Teukolsky S.~A.,  1972, \mn@doi [Astrophys. J.]
  {10.1086/151796}, 178, 347

\bibitem[\protect\citeauthoryear{Barnes \& Kasen}{Barnes \&
  Kasen}{2013}]{Barnes:2013wka}
Barnes J.,  Kasen D.,  2013, \mn@doi [Astrophys. J.]
  {10.1088/0004-637X/775/1/18}, 775, 18

\bibitem[\protect\citeauthoryear{Barnes, Kasen, Wu  \&
  Mart\'\i{}nez-Pinedo}{Barnes et~al.}{2016}]{Barnes:2016umi}
Barnes J.,  Kasen D.,  Wu M.-R.,   Mart\'\i{}nez-Pinedo G.,  2016, \mn@doi
  [Astrophys. J.] {10.3847/0004-637X/829/2/110}, 829, 110

\bibitem[\protect\citeauthoryear{Bartos \& Marka}{Bartos \&
  Marka}{2019}]{Bartos:2019cec}
Bartos I.,  Marka S.,  2019, \mn@doi [Nature] {10.1038/s41586-019-1113-7}, 569,
  85

\bibitem[\protect\citeauthoryear{Beniamini \& Piran}{Beniamini \&
  Piran}{2019}]{Beniamini:2019iop}
Beniamini P.,  Piran T.,  2019, \mn@doi [Mon. Not. Roy. Astron. Soc.]
  {10.1093/mnras/stz1589}, 487, 4847

\bibitem[\protect\citeauthoryear{Beniamini, Hotokezaka  \& Piran}{Beniamini
  et~al.}{2016}]{Beniamini:2016hoq}
Beniamini P.,  Hotokezaka K.,   Piran T.,  2016, \mn@doi [Astrophys. J. Lett.]
  {10.3847/2041-8205/829/1/L13}, 829, L13

\bibitem[\protect\citeauthoryear{Beniamini, Hotokezaka, van~der Horst  \&
  Kouveliotou}{Beniamini et~al.}{2019}]{Beniamini:2019bga}
Beniamini P.,  Hotokezaka K.,  van~der Horst A.,   Kouveliotou C.,  2019,
  \mn@doi [Mon. Not. Roy. Astron. Soc.] {10.1093/mnras/stz1391}, 487, 1426

\bibitem[\protect\citeauthoryear{Bhattacharya, Kumar  \& Smoot}{Bhattacharya
  et~al.}{2019}]{Bhattacharya:2018lmw}
Bhattacharya M.,  Kumar P.,   Smoot G.,  2019, \mn@doi [Mon. Not. Roy. Astron.
  Soc.] {10.1093/mnras/stz1147}, 486, 5289

\bibitem[\protect\citeauthoryear{Bhattacharya, Horiuchi  \&
  Murase}{Bhattacharya et~al.}{2022}]{Bhattacharya:2021cjc}
Bhattacharya M.,  Horiuchi S.,   Murase K.,  2022, \mn@doi [Mon. Not. Roy.
  Astron. Soc.] {10.1093/mnras/stac1721}, 514, 6011

\bibitem[\protect\citeauthoryear{Bhattacharya, Carpio, Murase  \&
  Horiuchi}{Bhattacharya et~al.}{2023}]{Bhattacharya:2022btx}
Bhattacharya M.,  Carpio J.~A.,  Murase K.,   Horiuchi S.,  2023, \mn@doi [Mon.
  Not. Roy. Astron. Soc.] {10.1093/mnras/stad494}, 521, 2391

\bibitem[\protect\citeauthoryear{Biscoveanu, Landry  \& Vitale}{Biscoveanu
  et~al.}{2022}]{Biscoveanu:2022iue}
Biscoveanu S.,  Landry P.,   Vitale S.,  2022, \mn@doi [Mon. Not. Roy. Astron.
  Soc.] {10.1093/mnras/stac3052}, 518, 5298

\bibitem[\protect\citeauthoryear{Bliss, Witt, Arcones, Montes  \&
  Pereira}{Bliss et~al.}{2018}]{Bliss:2018nhk}
Bliss J.,  Witt M.,  Arcones A.,  Montes F.,   Pereira J.,  2018, \mn@doi
  [Astrophys. J.] {10.3847/1538-4357/aaadbe}, 855, 135

\bibitem[\protect\citeauthoryear{Bovard, Martin, Guercilena, Arcones, Rezzolla
  \& Korobkin}{Bovard et~al.}{2017}]{Bovard:2017mvn}
Bovard L.,  Martin D.,  Guercilena F.,  Arcones A.,  Rezzolla L.,   Korobkin
  O.,  2017, \mn@doi [Phys. Rev. D] {10.1103/PhysRevD.96.124005}, 96, 124005

\bibitem[\protect\citeauthoryear{Broekgaarden et~al.,}{Broekgaarden
  et~al.}{2021}]{Broekgaarden:2021iew}
Broekgaarden F.~S.,  et~al., 2021, \mn@doi [Mon. Not. Roy. Astron. Soc.]
  {10.1093/mnras/stab2716}, 508, 5028

\bibitem[\protect\citeauthoryear{Burbidge, Burbidge, Fowler  \& Hoyle}{Burbidge
  et~al.}{1957}]{b2fh}
Burbidge E.~M.,  Burbidge G.~R.,  Fowler W.~A.,   Hoyle F.,  1957, \mn@doi
  [Rev. Mod. Phys.] {10.1103/RevModPhys.29.547}, 29, 547

\bibitem[\protect\citeauthoryear{{Cain} et~al.,}{{Cain}
  et~al.}{2018}]{2018ApJ...864...43C}
{Cain} M.,  et~al., 2018, \mn@doi [\apj] {10.3847/1538-4357/aad37d}, \href
  {https://ui.adsabs.harvard.edu/abs/2018ApJ...864...43C} {864, 43}

\bibitem[\protect\citeauthoryear{Chen, Vitale  \& Foucart}{Chen
  et~al.}{2021}]{Chen:2021fro}
Chen H.-Y.,  Vitale S.,   Foucart F.,  2021, \mn@doi [Astrophys. J. Lett.]
  {10.3847/2041-8213/ac26c6}, 920, L3

\bibitem[\protect\citeauthoryear{Chen, Hu  \& Liang}{Chen
  et~al.}{2022}]{Chen:2022nsj}
Chen M.-H.,  Hu R.-C.,   Liang E.-W.,  2022, \mn@doi [Astrophys. J. Lett.]
  {10.3847/2041-8213/ac7470}, 932, L7

\bibitem[\protect\citeauthoryear{{Combi} \& {Siegel}}{{Combi} \&
  {Siegel}}{2023}]{combi2022}
{Combi} L.,  {Siegel} D.~M.,  2023, \mn@doi [\apj] {10.3847/1538-4357/acac29},
  \href {https://ui.adsabs.harvard.edu/abs/2023ApJ...944...28C} {944, 28}

\bibitem[\protect\citeauthoryear{C\^ot\'e, Belczynski, Fryer, Ritter, Paul,
  Wehmeyer  \& O'Shea}{C\^ot\'e et~al.}{2017}]{Cote:2016vla}
C\^ot\'e B.,  Belczynski K.,  Fryer C.~L.,  Ritter C.,  Paul A.,  Wehmeyer B.,
   O'Shea B.~W.,  2017, \mn@doi [Astrophys. J.] {10.3847/1538-4357/aa5c8d},
  836, 230

\bibitem[\protect\citeauthoryear{C\^ot\'e et~al.}{C\^ot\'e
  et~al.}{2018}]{Cote:2017evr}
C\^ot\'e B.,  et~al., 2018, \mn@doi [Astrophys. J.] {10.3847/1538-4357/aaad67},
  855, 99

\bibitem[\protect\citeauthoryear{C\^ot\'e et~al.}{C\^ot\'e
  et~al.}{2019}]{Cote:2018qku}
C\^ot\'e B.,  et~al., 2019, \mn@doi [Astrophys. J.] {10.3847/1538-4357/ab10db},
  875, 106

\bibitem[\protect\citeauthoryear{Coulter et~al.}{Coulter
  et~al.}{2017}]{Coulter:2017wya}
Coulter D.~A.,  et~al., 2017, \mn@doi [Science] {10.1126/science.aap9811}, 358,
  1556

\bibitem[\protect\citeauthoryear{Cowan, Sneden, Lawler, Aprahamian, Wiescher,
  Langanke, Mart\'\i{}nez-Pinedo  \& Thielemann}{Cowan
  et~al.}{2021}]{Cowan:2019pkx}
Cowan J.~J.,  Sneden C.,  Lawler J.~E.,  Aprahamian A.,  Wiescher M.,  Langanke
  K.,  Mart\'\i{}nez-Pinedo G.,   Thielemann F.-K.,  2021, \mn@doi [Rev. Mod.
  Phys.] {10.1103/RevModPhys.93.015002}, 93, 15002

\bibitem[\protect\citeauthoryear{Cybert et~al.,}{Cybert et~al.}{2010}]{reaclib}
Cybert R.~H.,  et~al., 2010, \mn@doi [The Astrophysical Journal Supplement
  Series] {10.1088/0067-0049/189/1/240}, 189, 240

\bibitem[\protect\citeauthoryear{Desai, Siegel  \& Metzger}{Desai
  et~al.}{2022}]{Desai:2022lyi}
Desai D.~K.,  Siegel D.~M.,   Metzger B.~D.,  2022, \mn@doi [Astrophys. J.]
  {10.3847/1538-4357/ac69da}, 931, 104

\bibitem[\protect\citeauthoryear{Dominik, Belczynski, Fryer, Holz, Berti,
  Bulik, Mandel  \& O'Shaughnessy}{Dominik et~al.}{2012}]{Dominik:2012kk}
Dominik M.,  Belczynski K.,  Fryer C.,  Holz D.,  Berti E.,  Bulik T.,  Mandel
  I.,   O'Shaughnessy R.,  2012, \mn@doi [Astrophys. J.]
  {10.1088/0004-637X/759/1/52}, 759, 52

\bibitem[\protect\citeauthoryear{Drozda, Belczynski, O'Shaughnessy, Bulik  \&
  Fryer}{Drozda et~al.}{2022}]{Drozda:2020qab}
Drozda P.,  Belczynski K.,  O'Shaughnessy R.,  Bulik T.,   Fryer C.~L.,  2022,
  \mn@doi [Astron. Astrophys.] {10.1051/0004-6361/202039418}, 667, A126

\bibitem[\protect\citeauthoryear{{Duncan} \& {Thompson}}{{Duncan} \&
  {Thompson}}{1992}]{1992ApJ...392L...9D}
{Duncan} R.~C.,  {Thompson} C.,  1992, \mn@doi [\apjl] {10.1086/186413}, \href
  {https://ui.adsabs.harvard.edu/abs/1992ApJ...392L...9D} {392, L9}

\bibitem[\protect\citeauthoryear{Ekanger, Bhattacharya  \& Horiuchi}{Ekanger
  et~al.}{2022}]{Ekanger:2022tia}
Ekanger N.,  Bhattacharya M.,   Horiuchi S.,  2022, \mn@doi [Mon. Not. Roy.
  Astron. Soc.] {10.1093/mnras/stac896}, 513, 405

\bibitem[\protect\citeauthoryear{Foucart, Hinderer  \& Nissanke}{Foucart
  et~al.}{2018}]{Foucart:2018rjc}
Foucart F.,  Hinderer T.,   Nissanke S.,  2018, \mn@doi [Phys. Rev. D]
  {10.1103/PhysRevD.98.081501}, 98, 081501

\bibitem[\protect\citeauthoryear{Fragione}{Fragione}{2021}]{Fragione:2021cvv}
Fragione G.,  2021, \mn@doi [Astrophys. J. Lett.] {10.3847/2041-8213/ac3bcd},
  923, L2

\bibitem[\protect\citeauthoryear{Fujibayashi, Shibata, Wanajo, Kiuchi, Kyutoku
  \& Sekiguchi}{Fujibayashi et~al.}{2020}]{Fujibayashi:2020jfr}
Fujibayashi S.,  Shibata M.,  Wanajo S.,  Kiuchi K.,  Kyutoku K.,   Sekiguchi
  Y.,  2020, \mn@doi [Phys. Rev. D] {10.1103/PhysRevD.102.123014}, 102, 123014

\bibitem[\protect\citeauthoryear{{Fujibayashi}, {Kiuchi}, {Wanajo}, {Kyutoku},
  {Sekiguchi}  \& {Shibata}}{{Fujibayashi} et~al.}{2023}]{fujibayashi2022}
{Fujibayashi} S.,  {Kiuchi} K.,  {Wanajo} S.,  {Kyutoku} K.,  {Sekiguchi} Y.,
  {Shibata} M.,  2023, \mn@doi [\apj] {10.3847/1538-4357/ac9ce0}, \href
  {https://ui.adsabs.harvard.edu/abs/2023ApJ...942...39F} {942, 39}

\bibitem[\protect\citeauthoryear{Giacobbo \& Mapelli}{Giacobbo \&
  Mapelli}{2018}]{Giacobbo:2018etu}
Giacobbo N.,  Mapelli M.,  2018, \mn@doi [Mon. Not. Roy. Astron. Soc.]
  {10.1093/mnras/sty1999}, 480, 2011

\bibitem[\protect\citeauthoryear{Goldstein et~al.}{Goldstein
  et~al.}{2017}]{Goldstein:2017mmi}
Goldstein A.,  et~al., 2017, \mn@doi [Astrophys. J. Lett.]
  {10.3847/2041-8213/aa8f41}, 848, L14

\bibitem[\protect\citeauthoryear{Gomez, Berger, Nicholl, Blanchard  \&
  Hosseinzadeh}{Gomez et~al.}{2022}]{Gomez:2022mdw}
Gomez S.,  Berger E.,  Nicholl M.,  Blanchard P.~K.,   Hosseinzadeh G.,  2022,
  \mn@doi [Astrophys. J.] {10.3847/1538-4357/ac9842}, 941, 107

\bibitem[\protect\citeauthoryear{Goriely \& Janka}{Goriely \&
  Janka}{2016}]{Goriely:2016gfe}
Goriely S.,  Janka H.~T.,  2016, \mn@doi [Mon. Not. Roy. Astron. Soc.]
  {10.1093/mnras/stw946}, 459, 4174

\bibitem[\protect\citeauthoryear{{Gupta}, {Borhanian}, {Dhani},
  {Chattopadhyay}, {Kashyap}, {Villar}  \& {Sathyaprakash}}{{Gupta}
  et~al.}{2023}]{Gupta:2023evt}
{Gupta} I.,  {Borhanian} S.,  {Dhani} A.,  {Chattopadhyay} D.,  {Kashyap} R.,
  {Villar} V.~A.,   {Sathyaprakash} B.~S.,  2023, \mn@doi [arXiv e-prints]
  {10.48550/arXiv.2301.08763}, \href
  {https://ui.adsabs.harvard.edu/abs/2023arXiv230108763G} {p. arXiv:2301.08763}

\bibitem[\protect\citeauthoryear{Hajela et~al.}{Hajela
  et~al.}{2022}]{Hajela:2021faz}
Hajela A.,  et~al., 2022, \mn@doi [Astrophys. J. Lett.]
  {10.3847/2041-8213/ac504a}, 927, L17

\bibitem[\protect\citeauthoryear{Halevi \& M\"osta}{Halevi \&
  M\"osta}{2018}]{Halevi:2018vgp}
Halevi G.,  M\"osta P.,  2018, \mn@doi [Mon. Not. Roy. Astron. Soc.]
  {10.1093/mnras/sty797}, 477, 2366

\bibitem[\protect\citeauthoryear{{Hansen} et~al.,}{{Hansen}
  et~al.}{2012}]{2012A&A...545A..31H}
{Hansen} C.~J.,  et~al., 2012, \mn@doi [\aap] {10.1051/0004-6361/201118643},
  \href {https://ui.adsabs.harvard.edu/abs/2012A&A...545A..31H} {545, A31}

\bibitem[\protect\citeauthoryear{Hayashi, Fujibayashi, Kiuchi, Kyutoku,
  Sekiguchi  \& Shibata}{Hayashi et~al.}{2022}]{Hayashi:2021oxy}
Hayashi K.,  Fujibayashi S.,  Kiuchi K.,  Kyutoku K.,  Sekiguchi Y.,   Shibata
  M.,  2022, \mn@doi [Phys. Rev. D] {10.1103/PhysRevD.106.023008}, 106, 023008

\bibitem[\protect\citeauthoryear{{Henkel}, {Foucart}, {Raaijmakers}  \&
  {Nissanke}}{{Henkel} et~al.}{2022}]{Henkel:2022naw}
{Henkel} A.,  {Foucart} F.,  {Raaijmakers} G.,   {Nissanke} S.,  2022, \mn@doi
  [arXiv e-prints] {10.48550/arXiv.2207.07658}, \href
  {https://ui.adsabs.harvard.edu/abs/2022arXiv220707658H} {p. arXiv:2207.07658}

\bibitem[\protect\citeauthoryear{{Holmbeck}, {Surman}, {Roederer}, {McLaughlin}
   \& {Frebel}}{{Holmbeck} et~al.}{2022}]{Holmbeck:2022mog}
{Holmbeck} E.~M.,  {Surman} R.,  {Roederer} I.~U.,  {McLaughlin} G.~C.,
  {Frebel} A.,  2022, \mn@doi [arXiv e-prints] {10.48550/arXiv.2210.10122},
  \href {https://ui.adsabs.harvard.edu/abs/2022arXiv221010122H} {p.
  arXiv:2210.10122}

\bibitem[\protect\citeauthoryear{Honda, Aoki, Ishimaru, Wanajo  \& Ryan}{Honda
  et~al.}{2006}]{Honda:2006kp}
Honda S.,  Aoki W.,  Ishimaru Y.,  Wanajo S.,   Ryan S.~G.,  2006, \mn@doi
  [Astrophys. J.] {10.1086/503195}, 643, 1180

\bibitem[\protect\citeauthoryear{Horowitz et~al.}{Horowitz
  et~al.}{2019}]{Horowitz:2018ndv}
Horowitz C.~J.,  et~al., 2019, \mn@doi [J. Phys. G] {10.1088/1361-6471/ab0849},
  46, 083001

\bibitem[\protect\citeauthoryear{Hotokezaka, Piran  \& Paul}{Hotokezaka
  et~al.}{2015}]{Hotokezaka:2015zea}
Hotokezaka K.,  Piran T.,   Paul M.,  2015, \mn@doi [Nature Phys.]
  {10.1038/nphys3574}, 11, 1042

\bibitem[\protect\citeauthoryear{Hotokezaka, Beniamini  \& Piran}{Hotokezaka
  et~al.}{2018}]{Hotokezaka:2018aui}
Hotokezaka K.,  Beniamini P.,   Piran T.,  2018, \mn@doi [Int. J. Mod. Phys. D]
  {10.1142/S0218271818420051}, 27, 1842005

\bibitem[\protect\citeauthoryear{Ivezi\'c et~al.}{Ivezi\'c
  et~al.}{2019}]{LSST:2008ijt}
Ivezi\'c v.,  et~al., 2019, \mn@doi [Astrophys. J.] {10.3847/1538-4357/ab042c},
  873, 111

\bibitem[\protect\citeauthoryear{{Ji}, {Drout}  \& {Hansen}}{{Ji}
  et~al.}{2019}]{ji2019ApJ...882...40J}
{Ji} A.~P.,  {Drout} M.~R.,   {Hansen} T.~T.,  2019, \mn@doi [\apj]
  {10.3847/1538-4357/ab3291}, \href
  {https://ui.adsabs.harvard.edu/abs/2019ApJ...882...40J} {882, 40}

\bibitem[\protect\citeauthoryear{{Just}, {Bauswein}, {Ardevol Pulpillo},
  {Goriely}  \& {Janka}}{{Just} et~al.}{2015}]{2015just}
{Just} O.,  {Bauswein} A.,  {Ardevol Pulpillo} R.,  {Goriely} S.,   {Janka}
  H.~T.,  2015, \mn@doi [\mnras] {10.1093/mnras/stv009}, \href
  {https://ui.adsabs.harvard.edu/abs/2015MNRAS.448..541J} {448, 541}

\bibitem[\protect\citeauthoryear{Kajino, Aoki, Balantekin, Diehl, Famiano  \&
  Mathews}{Kajino et~al.}{2019}]{Kajino:2019abv}
Kajino T.,  Aoki W.,  Balantekin A.~B.,  Diehl R.,  Famiano M.~A.,   Mathews
  G.~J.,  2019, \mn@doi [Prog. Part. Nucl. Phys.] {10.1016/j.ppnp.2019.02.008},
  107, 109

\bibitem[\protect\citeauthoryear{Kasen, Fernandez  \& Metzger}{Kasen
  et~al.}{2015}]{Kasen:2014toa}
Kasen D.,  Fernandez R.,   Metzger B.,  2015, \mn@doi [Mon. Not. Roy. Astron.
  Soc.] {10.1093/mnras/stv721}, 450, 1777

\bibitem[\protect\citeauthoryear{Kasen, Metzger, Barnes, Quataert  \&
  Ramirez-Ruiz}{Kasen et~al.}{2017}]{Kasen:2017sxr}
Kasen D.,  Metzger B.,  Barnes J.,  Quataert E.,   Ramirez-Ruiz E.,  2017,
  \mn@doi [Nature] {10.1038/nature24453}, 551, 80

\bibitem[\protect\citeauthoryear{Kashiyama, Murase, Bartos, Kiuchi  \&
  Margutti}{Kashiyama et~al.}{2016}]{Kashiyama:2015eua}
Kashiyama K.,  Murase K.,  Bartos I.,  Kiuchi K.,   Margutti R.,  2016, \mn@doi
  [Astrophys. J.] {10.3847/0004-637X/818/1/94}, 818, 94

\bibitem[\protect\citeauthoryear{Kathirgamaraju, Giannios  \&
  Beniamini}{Kathirgamaraju et~al.}{2019}]{Kathirgamaraju:2019xwu}
Kathirgamaraju A.,  Giannios D.,   Beniamini P.,  2019, \mn@doi [Mon. Not. Roy.
  Astron. Soc.] {10.1093/mnras/stz1564}, 487, 3914

\bibitem[\protect\citeauthoryear{Kawaguchi, Kyutoku, Shibata  \&
  Tanaka}{Kawaguchi et~al.}{2016}]{Kawaguchi:2016ana}
Kawaguchi K.,  Kyutoku K.,  Shibata M.,   Tanaka M.,  2016, \mn@doi [Astrophys.
  J.] {10.3847/0004-637X/825/1/52}, 825, 52

\bibitem[\protect\citeauthoryear{Kim, Kalogera  \& Lorimer}{Kim
  et~al.}{2010}]{Kim:2006fm}
Kim C.,  Kalogera V.,   Lorimer D.~R.,  2010, \mn@doi [New Astron. Rev.]
  {10.1016/j.newar.2010.09.010}, 54, 148

\bibitem[\protect\citeauthoryear{{Kiuchi}, {Fujibayashi}, {Hayashi}, {Kyutoku},
  {Sekiguchi}  \& {Shibata}}{{Kiuchi} et~al.}{2022}]{Kiuchi:2022nin}
{Kiuchi} K.,  {Fujibayashi} S.,  {Hayashi} K.,  {Kyutoku} K.,  {Sekiguchi} Y.,
   {Shibata} M.,  2022, \mn@doi [arXiv e-prints] {10.48550/arXiv.2211.07637},
  \href {https://ui.adsabs.harvard.edu/abs/2022arXiv221107637K} {p.
  arXiv:2211.07637}

\bibitem[\protect\citeauthoryear{Kobayashi, Karakas  \& Lugaro}{Kobayashi
  et~al.}{2020}]{Kobayashi:2020jes}
Kobayashi C.,  Karakas A.~I.,   Lugaro M.,  2020, \mn@doi [Astrophys. J.]
  {10.3847/1538-4357/abae65}, 900, 179

\bibitem[\protect\citeauthoryear{{Kobayashi} et~al.,}{{Kobayashi}
  et~al.}{2023}]{Kobayashi:2022qlk}
{Kobayashi} C.,  et~al., 2023, \mn@doi [\apjl] {10.3847/2041-8213/acad82},
  \href {https://ui.adsabs.harvard.edu/abs/2023ApJ...943L..12K} {943, L12}

\bibitem[\protect\citeauthoryear{Kohri, Narayan  \& Piran}{Kohri
  et~al.}{2005}]{Kohri:2005tq}
Kohri K.,  Narayan R.,   Piran T.,  2005, \mn@doi [Astrophys. J.]
  {10.1086/431354}, 629, 341

\bibitem[\protect\citeauthoryear{Korobkin, Rosswog, Arcones  \&
  Winteler}{Korobkin et~al.}{2012}]{Korobkin:2012uy}
Korobkin O.,  Rosswog S.,  Arcones A.,   Winteler C.,  2012, \mn@doi [Mon. Not.
  Roy. Astron. Soc.] {10.1111/j.1365-2966.2012.21859.x}, 426, 1940

\bibitem[\protect\citeauthoryear{Kr\"uger \& Foucart}{Kr\"uger \&
  Foucart}{2020}]{Kruger:2020gig}
Kr\"uger C.~J.,  Foucart F.,  2020, \mn@doi [Phys. Rev. D]
  {10.1103/PhysRevD.101.103002}, 101, 103002

\bibitem[\protect\citeauthoryear{{Kullmann}, {Goriely}, {Just}, {Bauswein}  \&
  {Janka}}{{Kullmann} et~al.}{2022a}]{Kullmann:2022xuj}
{Kullmann} I.,  {Goriely} S.,  {Just} O.,  {Bauswein} A.,   {Janka} H.~T.,
  2022a, \mn@doi [arXiv e-prints] {10.48550/arXiv.2207.07421}, \href
  {https://ui.adsabs.harvard.edu/abs/2022arXiv220707421K} {p. arXiv:2207.07421}

\bibitem[\protect\citeauthoryear{Kullmann, Goriely, Just, Ardevol-Pulpillo,
  Bauswein  \& Janka}{Kullmann et~al.}{2022b}]{Kullmann:2021gvo}
Kullmann I.,  Goriely S.,  Just O.,  Ardevol-Pulpillo R.,  Bauswein A.,   Janka
  H.~T.,  2022b, \mn@doi [Mon. Not. Roy. Astron. Soc.]
  {10.1093/mnras/stab3393}, 510, 2804

\bibitem[\protect\citeauthoryear{Kyutoku, Shibata  \& Taniguchi}{Kyutoku
  et~al.}{2021}]{Kyutoku:2021icp}
Kyutoku K.,  Shibata M.,   Taniguchi K.,  2021, \mn@doi [Living Rev. Rel.]
  {10.1007/s41114-021-00033-4}, 24, 5

\bibitem[\protect\citeauthoryear{{LSST Science Collaboration} et~al.,}{{LSST
  Science Collaboration} et~al.}{2009}]{LSSTScience:2009jmu}
{LSST Science Collaboration} et~al., 2009, \mn@doi [arXiv e-prints]
  {10.48550/arXiv.0912.0201}, \href
  {https://ui.adsabs.harvard.edu/abs/2009arXiv0912.0201L} {p. arXiv:0912.0201}

\bibitem[\protect\citeauthoryear{Lackey, Nayyar  \& Owen}{Lackey
  et~al.}{2006}]{PhysRevD.73.024021}
Lackey B.~D.,  Nayyar M.,   Owen B.~J.,  2006, \mn@doi [Phys. Rev. D]
  {10.1103/PhysRevD.73.024021}, 73, 024021

\bibitem[\protect\citeauthoryear{{Lewin} \& {van der Klis}}{{Lewin} \& {van der
  Klis}}{2006}]{lewin2006}
{Lewin} W. H.~G.,  {van der Klis} M.,  2006, Compact Stellar X-ray Sources.
Cambridge Astrophysics, Cambridge University Press,
  \mn@doi{10.1017/CBO9780511536281}

\bibitem[\protect\citeauthoryear{Li \& Paczynski}{Li \&
  Paczynski}{1998}]{Li:1998bw}
Li L.-X.,  Paczynski B.,  1998, \mn@doi [Astrophys. J. Lett.] {10.1086/311680},
  507, L59

\bibitem[\protect\citeauthoryear{Lippuner \& Roberts}{Lippuner \&
  Roberts}{2015}]{Lippuner:2015gwa}
Lippuner J.,  Roberts L.~F.,  2015, \mn@doi [Astrophys. J.]
  {10.1088/0004-637X/815/2/82}, 815, 82

\bibitem[\protect\citeauthoryear{Lippuner \& Roberts}{Lippuner \&
  Roberts}{2017}]{Lippuner_2017}
Lippuner J.,  Roberts L.~F.,  2017, \mn@doi [The Astrophysical Journal
  Supplement Series] {10.3847/1538-4365/aa94cb}, 233, 18

\bibitem[\protect\citeauthoryear{{Lund}, {Engel}, {McLaughlin}, {Mumpower},
  {Ney}  \& {Surman}}{{Lund} et~al.}{2023}]{Lund:2022bsr}
{Lund} K.~A.,  {Engel} J.,  {McLaughlin} G.~C.,  {Mumpower} M.~R.,  {Ney}
  E.~M.,   {Surman} R.,  2023, \mn@doi [\apj] {10.3847/1538-4357/acaf56}, \href
  {https://ui.adsabs.harvard.edu/abs/2023ApJ...944..144L} {944, 144}

\bibitem[\protect\citeauthoryear{MacFadyen \& Woosley}{MacFadyen \&
  Woosley}{1999}]{MacFadyen:1998vz}
MacFadyen A.,  Woosley S.~E.,  1999, \mn@doi [Astrophys. J.] {10.1086/307790},
  524, 262

\bibitem[\protect\citeauthoryear{Mennekens \& Vanbeveren}{Mennekens \&
  Vanbeveren}{2016}]{Mennekens:2016jcs}
Mennekens N.,  Vanbeveren D.,  2016, \mn@doi [Astron. Astrophys.]
  {10.1051/0004-6361/201628193}, 589, A64

\bibitem[\protect\citeauthoryear{Metzger et~al.,}{Metzger
  et~al.}{2010}]{Metzger:2010sy}
Metzger B.~D.,  et~al., 2010, \mn@doi [Mon. Not. Roy. Astron. Soc.]
  {10.1111/j.1365-2966.2010.16864.x}, 406, 2650

\bibitem[\protect\citeauthoryear{Metzger, Giannios, Thompson, Bucciantini  \&
  Quataert}{Metzger et~al.}{2011}]{Metzger:2010pp}
Metzger B.~D.,  Giannios D.,  Thompson T.~A.,  Bucciantini N.,   Quataert E.,
  2011, \mn@doi [Mon. Not. Roy. Astron. Soc.]
  {10.1111/j.1365-2966.2011.18280.x}, 413, 2031

\bibitem[\protect\citeauthoryear{Mumpower, Kawano, Ullmann, Krtička  \&
  Sprouse}{Mumpower et~al.}{2017}]{Mumpower_2017}
Mumpower M.~R.,  Kawano T.,  Ullmann J.~L.,  Krtička M.,   Sprouse T.~M.,
  2017, \mn@doi [Physical Review C] {10.1103/physrevc.96.024612}, 96, 024612

\bibitem[\protect\citeauthoryear{Nedora, Radice, Bernuzzi, Perego, Daszuta,
  Endrizzi, Prakash  \& Schianchi}{Nedora et~al.}{2021}]{Nedora:2021eoj}
Nedora V.,  Radice D.,  Bernuzzi S.,  Perego A.,  Daszuta B.,  Endrizzi A.,
  Prakash A.,   Schianchi F.,  2021, \mn@doi [Mon. Not. Roy. Astron. Soc.]
  {10.1093/mnras/stab2004}, 506, 5908

\bibitem[\protect\citeauthoryear{Nedora, Dietrich, Shibata, Pohl  \&
  Crosato Menegazzi}{Nedora et~al.}{2023a}]{nedora2023}
Nedora V.,  Dietrich T.,  Shibata M.,  Pohl M.,   Crosato Menegazzi L.,
  2023a, \mn@doi [Monthly Notices of the Royal Astronomical Society]
  {10.1093/mnras/stad175}, 520, 2727

\bibitem[\protect\citeauthoryear{Nedora, Dietrich  \& Shibata}{Nedora
  et~al.}{2023b}]{Nedora:2023hiz}
Nedora V.,  Dietrich T.,   Shibata M.,  2023b, \mn@doi [Monthly Notices of the
  Royal Astronomical Society] {10.1093/mnras/stad2128}, 524, 5514

\bibitem[\protect\citeauthoryear{Nishimura, Takiwaki  \& Thielemann}{Nishimura
  et~al.}{2015}]{Nishimura:2015nca}
Nishimura N.,  Takiwaki T.,   Thielemann F.~K.,  2015, \mn@doi [Astrophys. J.]
  {10.1088/0004-637X/810/2/109}, 810, 109

\bibitem[\protect\citeauthoryear{Pian et~al.}{Pian et~al.}{2017}]{Pian:2017gtc}
Pian E.,  et~al., 2017, \mn@doi [Nature] {10.1038/nature24298}, 551, 67

\bibitem[\protect\citeauthoryear{Pruet, Thompson  \& Hoffman}{Pruet
  et~al.}{2004}]{Pruet:2003ts}
Pruet J.,  Thompson T.,   Hoffman R.~D.,  2004, \mn@doi [Astrophys. J.]
  {10.1086/382036}, 606, 1006

\bibitem[\protect\citeauthoryear{Psaltis, Arcones, Montes, Mohr, Hansen, Jacobi
   \& Schatz}{Psaltis et~al.}{2022}]{Psaltis:2022jgr}
Psaltis A.,  Arcones A.,  Montes F.,  Mohr P.,  Hansen C.~J.,  Jacobi M.,
  Schatz H.,  2022, \mn@doi [Astrophys. J.] {10.3847/1538-4357/ac7da7}, 935, 27

\bibitem[\protect\citeauthoryear{Qian \& Woosley}{Qian \&
  Woosley}{1996}]{Qian:1996xt}
Qian Y.~Z.,  Woosley S.~E.,  1996, \mn@doi [Astrophys. J.] {10.1086/177973},
  471, 331

\bibitem[\protect\citeauthoryear{Raaijmakers et~al.}{Raaijmakers
  et~al.}{2021}]{Raaijmakers:2021slr}
Raaijmakers G.,  et~al., 2021, \mn@doi [Astrophys. J.]
  {10.3847/1538-4357/ac222d}, 922, 269

\bibitem[\protect\citeauthoryear{Radice, Perego, Hotokezaka, Fromm, Bernuzzi
  \& Roberts}{Radice et~al.}{2018}]{Radice:2018pdn}
Radice D.,  Perego A.,  Hotokezaka K.,  Fromm S.~A.,  Bernuzzi S.,   Roberts
  L.~F.,  2018, \mn@doi [Astrophys. J.] {10.3847/1538-4357/aaf054}, 869, 130

\bibitem[\protect\citeauthoryear{Reichert, Obergaulinger, Eichler, Aloy  \&
  Arcones}{Reichert et~al.}{2021}]{Reichert:2020mjo}
Reichert M.,  Obergaulinger M.,  Eichler M.,  Aloy M.-A.,   Arcones A.,  2021,
  \mn@doi [Mon. Not. Roy. Astron. Soc.] {10.1093/mnras/stab029}, 501, 5733

\bibitem[\protect\citeauthoryear{{Reichert}, {Obergaulinger}, {Aloy}, {Gabler},
  {Arcones}  \& {Thielemann}}{{Reichert} et~al.}{2023}]{reichert2022}
{Reichert} M.,  {Obergaulinger} M.,  {Aloy} M.~{\'A}.,  {Gabler} M.,  {Arcones}
  A.,   {Thielemann} F.~K.,  2023, \mn@doi [\mnras] {10.1093/mnras/stac3185},
  \href {https://ui.adsabs.harvard.edu/abs/2023MNRAS.518.1557R} {518, 1557}

\bibitem[\protect\citeauthoryear{Roberts et~al.,}{Roberts
  et~al.}{2017}]{Roberts:2016igt}
Roberts L.~F.,  et~al., 2017, \mn@doi [Mon. Not. Roy. Astron. Soc.]
  {10.1093/mnras/stw2622}, 464, 3907

\bibitem[\protect\citeauthoryear{{Roederer} et~al.,}{{Roederer}
  et~al.}{2022a}]{HD222925}
{Roederer} I.~U.,  et~al., 2022a, \mn@doi [\apjs] {10.3847/1538-4365/ac5cbc},
  \href {https://ui.adsabs.harvard.edu/abs/2022ApJS..260...27R} {260, 27}

\bibitem[\protect\citeauthoryear{Roederer et~al.}{Roederer
  et~al.}{2022b}]{Roederer:2022exr}
Roederer I.~U.,  et~al., 2022b, \mn@doi [Astrophys. J.]
  {10.3847/1538-4357/ac85bc}, 936, 84

\bibitem[\protect\citeauthoryear{Rom\'an-Garza et~al.}{Rom\'an-Garza
  et~al.}{2021}]{Roman-Garza:2020uou}
Rom\'an-Garza J.,  et~al., 2021, \mn@doi [Astrophys. J. Lett.]
  {10.3847/2041-8213/abf42c}, 912, L23

\bibitem[\protect\citeauthoryear{Rosswog \& Korobkin}{Rosswog \&
  Korobkin}{2022}]{rosswog2022}
Rosswog S.,  Korobkin O.,  2022, Annalen Phys., 2022, 2200306

\bibitem[\protect\citeauthoryear{Rosswog, Feindt, Korobkin, Wu, Sollerman,
  Goobar  \& Martinez-Pinedo}{Rosswog et~al.}{2017}]{Rosswog:2016dhy}
Rosswog S.,  Feindt U.,  Korobkin O.,  Wu M.~R.,  Sollerman J.,  Goobar A.,
  Martinez-Pinedo G.,  2017, \mn@doi [Class. Quant. Grav.]
  {10.1088/1361-6382/aa68a9}, 34, 104001

\bibitem[\protect\citeauthoryear{Sadeh, Guttman  \& Waxman}{Sadeh
  et~al.}{2022}]{Sadeh:2022enp}
Sadeh G.,  Guttman O.,   Waxman E.,  2022, \mn@doi [Mon. Not. Roy. Astron.
  Soc.] {10.1093/mnras/stac3260}, 518, 2102

\bibitem[\protect\citeauthoryear{Santoliquido, Mapelli, Artale  \&
  Boco}{Santoliquido et~al.}{2022}]{Santoliquido:2022kyu}
Santoliquido F.,  Mapelli M.,  Artale M.~C.,   Boco L.,  2022, \mn@doi [Mon.
  Not. Roy. Astron. Soc.] {10.1093/mnras/stac2384}, 516, 3297

\bibitem[\protect\citeauthoryear{Savchenko et~al.}{Savchenko
  et~al.}{2017}]{Savchenko:2017ffs}
Savchenko V.,  et~al., 2017, \mn@doi [Astrophys. J. Lett.]
  {10.3847/2041-8213/aa8f94}, 848, L15

\bibitem[\protect\citeauthoryear{Siegel, Barnes  \& Metzger}{Siegel
  et~al.}{2019}]{Siegel:2018zxq}
Siegel D.~M.,  Barnes J.,   Metzger B.~D.,  2019, \mn@doi [Nature]
  {10.1038/s41586-019-1136-0}, 569, 241

\bibitem[\protect\citeauthoryear{Simon et~al.}{Simon
  et~al.}{2023}]{Simon:2022prp}
Simon J.~D.,  et~al., 2023, \mn@doi [Astrophys. J.] {10.3847/1538-4357/aca9d1},
  944, 43

\bibitem[\protect\citeauthoryear{Smartt et~al.}{Smartt
  et~al.}{2017}]{Smartt:2017fuw}
Smartt S.~J.,  et~al., 2017, \mn@doi [Nature] {10.1038/nature24303}, 551, 75

\bibitem[\protect\citeauthoryear{{Sneden}, {Cowan}  \& {Gallino}}{{Sneden}
  et~al.}{2008}]{sneden2008ARA&A..46..241S}
{Sneden} C.,  {Cowan} J.~J.,   {Gallino} R.,  2008, \mn@doi [\araa]
  {10.1146/annurev.astro.46.060407.145207}, \href
  {https://ui.adsabs.harvard.edu/abs/2008ARA&A..46..241S} {46, 241}

\bibitem[\protect\citeauthoryear{Sneppen, Watson, Bauswein, Just, Kotak, Nakar,
  Poznanski  \& Sim}{Sneppen et~al.}{2023}]{Sneppen:2023vkk}
Sneppen A.,  Watson D.,  Bauswein A.,  Just O.,  Kotak R.,  Nakar E.,
  Poznanski D.,   Sim S.,  2023, \mn@doi [Nature] {10.1038/s41586-022-05616-x},
  614, 436

\bibitem[\protect\citeauthoryear{Soares-Santos et~al.}{Soares-Santos
  et~al.}{2017}]{DES:2017kbs}
Soares-Santos M.,  et~al., 2017, \mn@doi [Astrophys. J. Lett.]
  {10.3847/2041-8213/aa9059}, 848, L16

\bibitem[\protect\citeauthoryear{{Spite}, {Spite}, {Barbuy}, {Bonifacio},
  {Caffau}  \& {Fran{\c{c}}ois}}{{Spite} et~al.}{2018}]{2018A&A...611A..30S}
{Spite} F.,  {Spite} M.,  {Barbuy} B.,  {Bonifacio} P.,  {Caffau} E.,
  {Fran{\c{c}}ois} P.,  2018, \mn@doi [\aap] {10.1051/0004-6361/201732096},
  \href {https://ui.adsabs.harvard.edu/abs/2018A&A...611A..30S} {611, A30}

\bibitem[\protect\citeauthoryear{Takahashi, Witti  \& Janka}{Takahashi
  et~al.}{1994}]{Takahashi:1994yz}
Takahashi K.,  Witti J.,   Janka H.~T.,  1994, Astron. Astrophys., 286, 857

\bibitem[\protect\citeauthoryear{Tanaka et~al.}{Tanaka
  et~al.}{2018}]{Tanaka:2017lxb}
Tanaka M.,  et~al., 2018, \mn@doi [Astrophys. J.] {10.3847/1538-4357/aaa0cb},
  852, 109

\bibitem[\protect\citeauthoryear{Tanaka, Kato, Gaigalas  \& Kawaguchi}{Tanaka
  et~al.}{2020}]{Tanaka:2019iqp}
Tanaka M.,  Kato D.,  Gaigalas G.,   Kawaguchi K.,  2020, \mn@doi [Mon. Not.
  Roy. Astron. Soc.] {10.1093/mnras/staa1576}, 496, 1369

\bibitem[\protect\citeauthoryear{Tarumi, Hotokezaka  \& Beniamini}{Tarumi
  et~al.}{2021}]{Tarumi:2021xvw}
Tarumi Y.,  Hotokezaka K.,   Beniamini P.,  2021, \mn@doi [Astrophys. J. Lett.]
  {10.3847/2041-8213/abfe13}, 913, L30

\bibitem[\protect\citeauthoryear{Taylor et~al.}{Taylor
  et~al.}{2014}]{Taylor:2014rlo}
Taylor M.,  et~al., 2014, \mn@doi [Astrophys. J.]
  {10.1088/0004-637X/792/2/135}, 792, 135

\bibitem[\protect\citeauthoryear{{The LIGO Scientific Collaboration}
  et~al.,}{{The LIGO Scientific Collaboration} et~al.}{2021}]{ligo2111.03634}
{The LIGO Scientific Collaboration} et~al., 2021, \mn@doi [arXiv e-prints]
  {10.48550/arXiv.2111.03634}, \href
  {https://ui.adsabs.harvard.edu/abs/2021arXiv211103634T} {p. arXiv:2111.03634}

\bibitem[\protect\citeauthoryear{Thielemann, Eichler, Panov  \&
  Wehmeyer}{Thielemann et~al.}{2017}]{Thielemann:2017acv}
Thielemann F.~K.,  Eichler M.,  Panov I.~V.,   Wehmeyer B.,  2017, \mn@doi
  [Ann. Rev. Nucl. Part. Sci.] {10.1146/annurev-nucl-101916-123246}, 67, 253

\bibitem[\protect\citeauthoryear{Thompson, Chang  \& Quataert}{Thompson
  et~al.}{2004}]{Thompson_2004}
Thompson T.~A.,  Chang P.,   Quataert E.,  2004, \mn@doi [The Astrophysical
  Journal] {10.1086/421969}, 611, 380–393

\bibitem[\protect\citeauthoryear{Travaglio, Gallino, Arnone, Cowan, Jordan  \&
  Sneden}{Travaglio et~al.}{2004}]{Travaglio:2003qq}
Travaglio C.,  Gallino R.,  Arnone E.,  Cowan J.,  Jordan F.,   Sneden C.,
  2004, \mn@doi [Astrophys. J.] {10.1086/380507}, 601, 864

\bibitem[\protect\citeauthoryear{{Usov}}{{Usov}}{1992}]{1992Natur.357..472U}
{Usov} V.~V.,  1992, \mn@doi [\nat] {10.1038/357472a0}, \href
  {https://ui.adsabs.harvard.edu/abs/1992Natur.357..472U} {357, 472}

\bibitem[\protect\citeauthoryear{Valenti et~al.,}{Valenti
  et~al.}{2017}]{Valenti:2017ngx}
Valenti S.,  et~al., 2017, \mn@doi [Astrophys. J. Lett.]
  {10.3847/2041-8213/aa8edf}, 848, L24

\bibitem[\protect\citeauthoryear{Villar et~al.}{Villar
  et~al.}{2017}]{Villar:2017wcc}
Villar V.~A.,  et~al., 2017, \mn@doi [Astrophys. J. Lett.]
  {10.3847/2041-8213/aa9c84}, 851, L21

\bibitem[\protect\citeauthoryear{Vlasov, Metzger  \& Thompson}{Vlasov
  et~al.}{2014}]{Vlasov:2014ara}
Vlasov A.~D.,  Metzger B.~D.,   Thompson T.~A.,  2014, \mn@doi [Mon. Not. Roy.
  Astron. Soc.] {10.1093/mnras/stu1667}, 444, 3537

\bibitem[\protect\citeauthoryear{Vlasov, Metzger, Lippuner, Roberts  \&
  Thompson}{Vlasov et~al.}{2017}]{Vlasov:2017nou}
Vlasov A.~D.,  Metzger B.~D.,  Lippuner J.,  Roberts L.~F.,   Thompson T.~A.,
  2017, \mn@doi [Mon. Not. Roy. Astron. Soc.] {10.1093/mnras/stx478}, 468, 1522

\bibitem[\protect\citeauthoryear{Waxman, Ofek, Kushnir  \& Gal-Yam}{Waxman
  et~al.}{2018}]{Waxman:2017sqv}
Waxman E.,  Ofek E.~O.,  Kushnir D.,   Gal-Yam A.,  2018, \mn@doi [Mon. Not.
  Roy. Astron. Soc.] {10.1093/mnras/sty2441}, 481, 3423

\bibitem[\protect\citeauthoryear{Wheeler, Yi, Hoflich  \& Wang}{Wheeler
  et~al.}{2000}]{Wheeler_2000}
Wheeler J.~C.,  Yi I.,  Hoflich P.,   Wang L.,  2000, \mn@doi [The
  Astrophysical Journal] {10.1086/309055}, 537, 810–823

\bibitem[\protect\citeauthoryear{{Winteler}}{{Winteler}}{2014}]{2014PhDT.......206W}
{Winteler} C.,  2014, PhD thesis, University of Basel

\bibitem[\protect\citeauthoryear{Winteler, Kaeppeli, Perego, Arcones, Vasset,
  Nishimura, Liebendoerfer  \& Thielemann}{Winteler
  et~al.}{2012}]{Winteler:2012hu}
Winteler C.,  Kaeppeli R.,  Perego A.,  Arcones A.,  Vasset N.,  Nishimura N.,
  Liebendoerfer M.,   Thielemann F.-K.,  2012, \mn@doi [Astrophys. J. Lett.]
  {10.1088/2041-8205/750/1/L22}, 750, L22

\bibitem[\protect\citeauthoryear{Witt et~al.,}{Witt
  et~al.}{2021}]{Witt:2021gwk}
Witt M.,  et~al., 2021, \mn@doi [Astrophys. J.] {10.3847/1538-4357/ac1a6d},
  921, 19

\bibitem[\protect\citeauthoryear{{Wu}, {Wang}, {Shi}, {Zhao}  \& {Grupp}}{{Wu}
  et~al.}{2015}]{2015A&A...579A...8W}
{Wu} X.,  {Wang} L.,  {Shi} J.,  {Zhao} G.,   {Grupp} F.,  2015, \mn@doi [\aap]
  {10.1051/0004-6361/201525679}, \href
  {https://ui.adsabs.harvard.edu/abs/2015A&A...579A...8W} {579, A8}

\bibitem[\protect\citeauthoryear{Yong et~al.}{Yong et~al.}{2021}]{Yong:2021nkh}
Yong D.,  et~al., 2021, \mn@doi [Nature] {10.1038/s41586-021-03611-2}, 595, 223

\bibitem[\protect\citeauthoryear{Zhu, Yang, Liu, Huang, Zhang, Li, Yu  \&
  Gao}{Zhu et~al.}{2020}]{Zhu:2020inc}
Zhu J.-P.,  Yang Y.-P.,  Liu L.-D.,  Huang Y.,  Zhang B.,  Li Z.,  Yu Y.-W.,
  Gao H.,  2020, \mn@doi [Astrophys. J.] {10.3847/1538-4357/ab93bf}, 897, 20

\bibitem[\protect\citeauthoryear{Zhu, Lund, Barnes, Sprouse, Vassh, McLaughlin,
  Mumpower  \& Surman}{Zhu et~al.}{2021a}]{Zhu:2020eyk}
Zhu Y.~L.,  Lund K.,  Barnes J.,  Sprouse T.~M.,  Vassh N.,  McLaughlin G.~C.,
  Mumpower M.~R.,   Surman R.,  2021a, \mn@doi [Astrophys. J.]
  {10.3847/1538-4357/abc69e}, 906, 94

\bibitem[\protect\citeauthoryear{Zhu et~al.}{Zhu et~al.}{2021b}]{Zhu:2020ffa}
Zhu J.-P.,  et~al., 2021b, \mn@doi [Astrophys. J.] {10.3847/1538-4357/abfe5e},
  917, 24

\bibitem[\protect\citeauthoryear{Zhu, Yang, Zhang, Gao  \& Yu}{Zhu
  et~al.}{2022}]{Zhu:2021zmy}
Zhu J.-P.,  Yang Y.-P.,  Zhang B.,  Gao H.,   Yu Y.-W.,  2022, \mn@doi
  [Astrophys. J.] {10.3847/1538-4357/ac8e60}, 938, 147

\bibitem[\protect\citeauthoryear{Zhu et~al.}{Zhu et~al.}{2023}]{Zhu:2021ram}
Zhu J.-P.,  et~al., 2023, \mn@doi [Astrophys. J.] {10.3847/1538-4357/aca527},
  942, 88

\bibitem[\protect\citeauthoryear{{de Haas}, {Bosch}, {M{\"o}sta}, {Curtis}  \&
  {Schut}}{{de Haas} et~al.}{2022}]{deHaas2022}
{de Haas} S.,  {Bosch} P.,  {M{\"o}sta} P.,  {Curtis} S.,   {Schut} N.,  2022,
  \mn@doi [arXiv e-prints] {10.48550/arXiv.2208.05330}, \href
  {https://ui.adsabs.harvard.edu/abs/2022arXiv220805330D} {p. arXiv:2208.05330}

\makeatother
\end{thebibliography}

\appendix

\section{Detectability dependence on NS EOS}\label{eosappendix}

\begin{figure*}
\includegraphics[width=\linewidth]{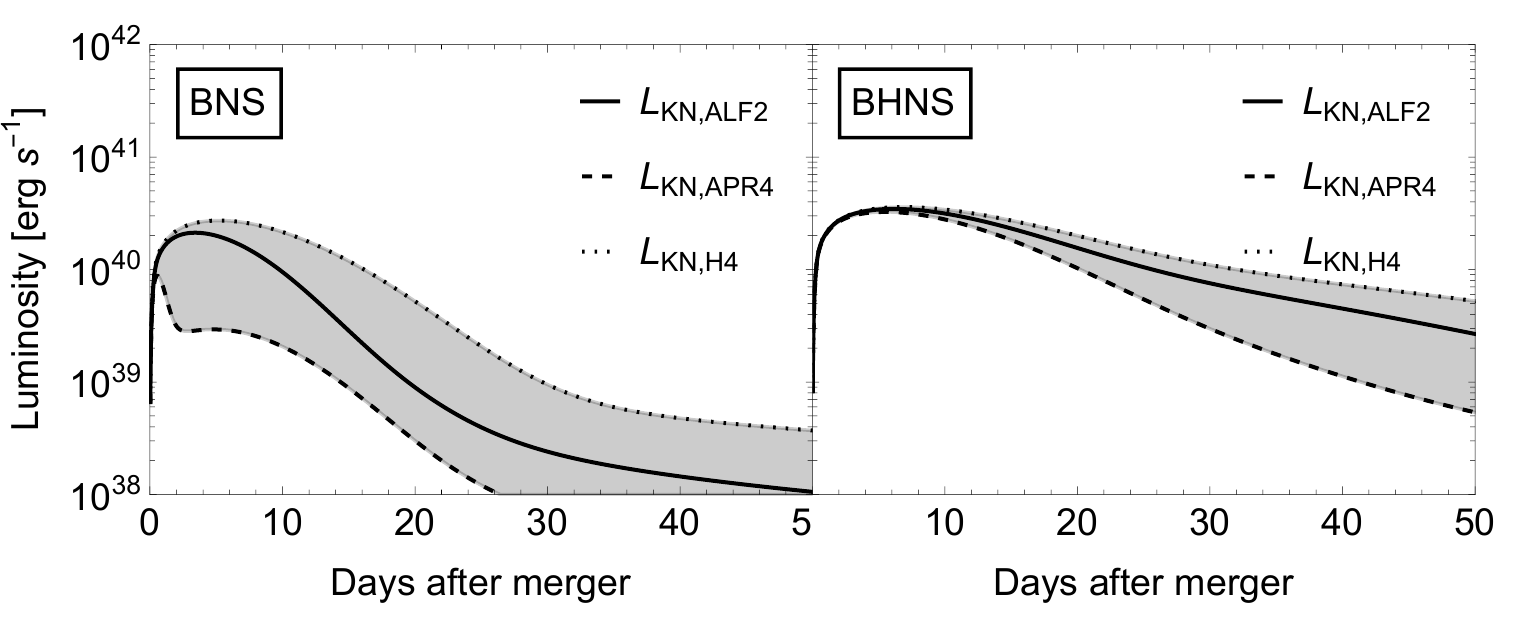}
\caption{Bolometric light curves for the BNS and BHNS kilonovae, varying the NS EOS. In both panels, the solid black line represents our fiducial ALF2 EOS, dashed black is the softer APR4 EOS, and dotted is the stiffer H4 EOS. The EOS makes a larger impact on the BNS merger kilonova light curve compared to BHNS mergers, as seen in the left-hand panel. Because the APR4 EOS is softer, the BNS wind does not eject as much mass with our fiducial parameters, and the resulting early time light curve is not very luminous. Besides this special case, the effect of the EOS is primarily to lengthen (for softer EOS) or shorten (for stiffer EOS) the duration of the signal. This effect is more apparent in the case of the BHNS kilonova shown in the right-hand panel.}
\label{fig:EOSknlightcurves}
\end{figure*}

In this section we quantify and discuss the effect that varying the EOS has on the BNS and BHNS merger observables. We perform the same calculations as before but with EOS-dependent compactness values for equations (\ref{bnsdynamical})-(\ref{bhnsdynejecta}), and (\ref{bhnsremnant}). This yields different values for the mass ejected in typical BNS/BHNS mergers and the corresponding thermalization fraction, equation (\ref{thermfrac}).

The resulting light curves can be seen in Fig.~\ref{fig:EOSknlightcurves}, where the solid-black line is the total kilonova luminosity for the ALF2 EOS (same as the black lines in Fig.~\ref{fig:knlightcurves}), the black-dashed line is for the softer APR4 EOS \citep{PhysRevC.56.2261}, and the black-dotted line is for the stiffer H4 EOS \citep{PhysRevD.73.024021}. The soft APR4 EOS plays the biggest role in BNS mergers (left-hand panel), where the initial luminosity is very low. If the NS compactness in BNS mergers is too high, as in the APR4 EOS, no accretion disc may form. Compactness values above 0.182 in \citet{Kruger:2020gig} result in little to no accretion disc being formed. Very little wind mass ejected leads to a very short and dim component, followed by a longer component from the dynamical ejecta. In the BHNS case (right-hand panel), the EOS primarily serves to lengthen (for stiff H4) or shorten (for soft APR4) the light curve in time, without changing the peak luminosity appreciably. This yields the following projected BNS rates at LSST (see $\S$\ref{prospects}): 3.1, 7.1, and 8.2 detections per year for the APR4, ALF2, and H4 EOS, respectively. The corresponding BHNS yields are 1.8, 1.8, and 1.9 detections per year for the APR4, ALF2, and H4 EOS, respectively. Finally, the EOS plays a large role in the fraction of BHNS mergers that successfully eject mass. This fraction is 52, 81, and 98 per cent for the APR4, ALF2, and H4 EOS, respectively.

\section{Mass ejected variation in mergers}\label{massejectedappendix}

Here, we show the variation in mass ejected for the BNS and BHNS merger scenarios. In the analysis presented in the main text, we assume the most probable initial parameters based on distributions theoretically found in nature. Depending on the initial NS and BH parameters, however, both the dynamical mass and wind mass ejected can vary by up to a factor of $\sim10$.

\begin{figure*}
\includegraphics[width=0.32\textwidth]{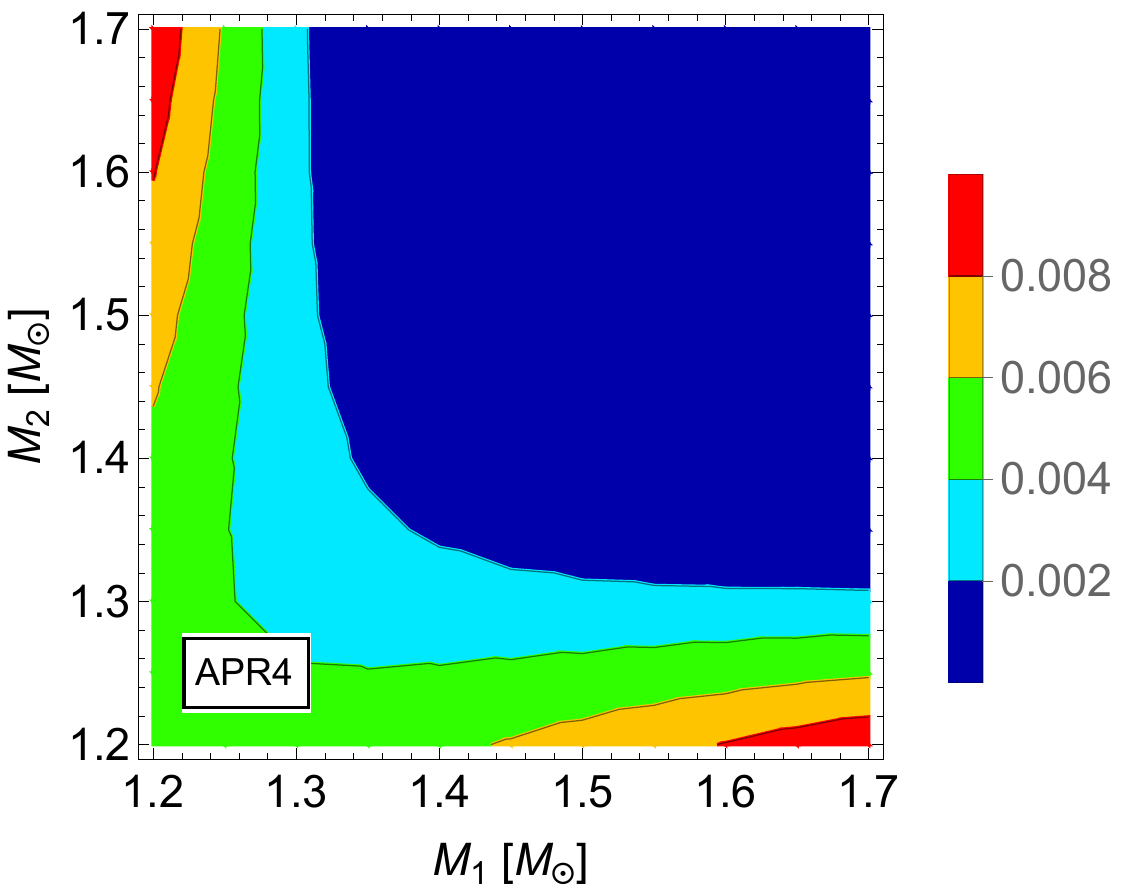}
\includegraphics[width=0.32\textwidth]{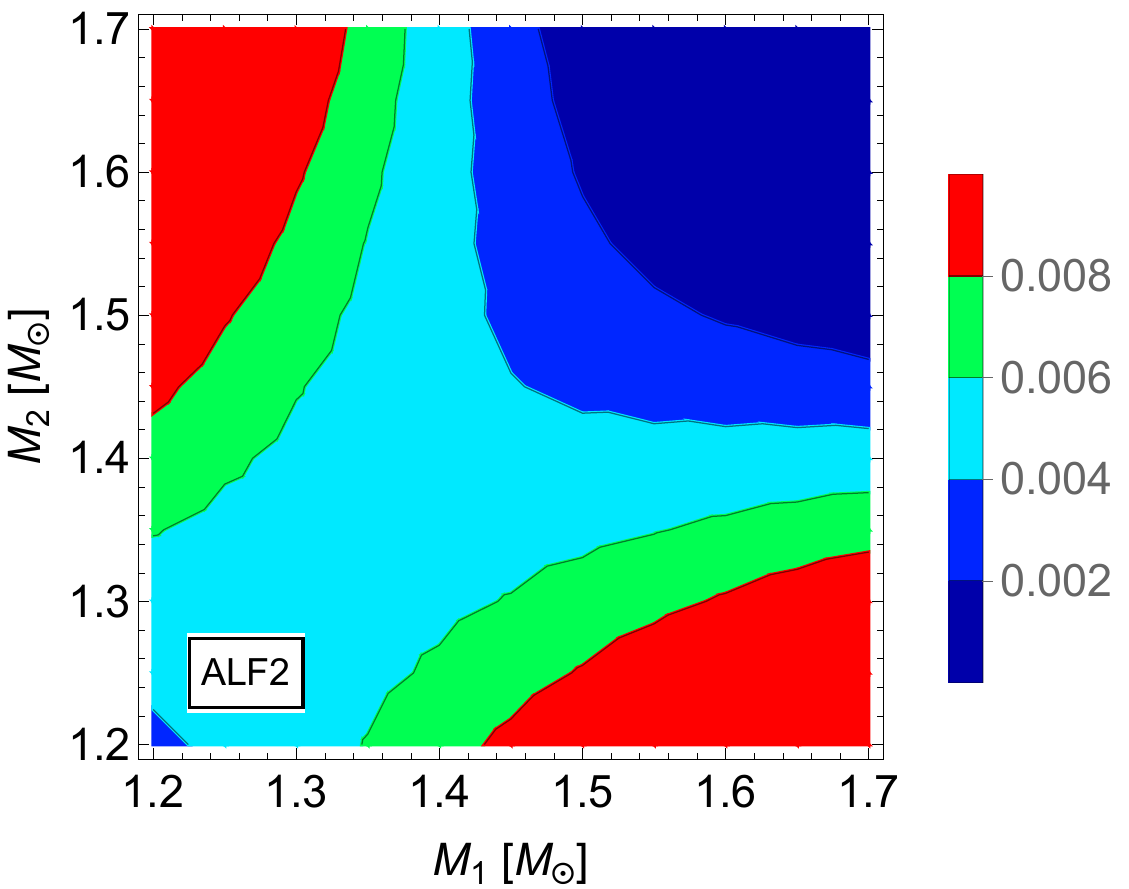}
\includegraphics[width=0.32\textwidth]{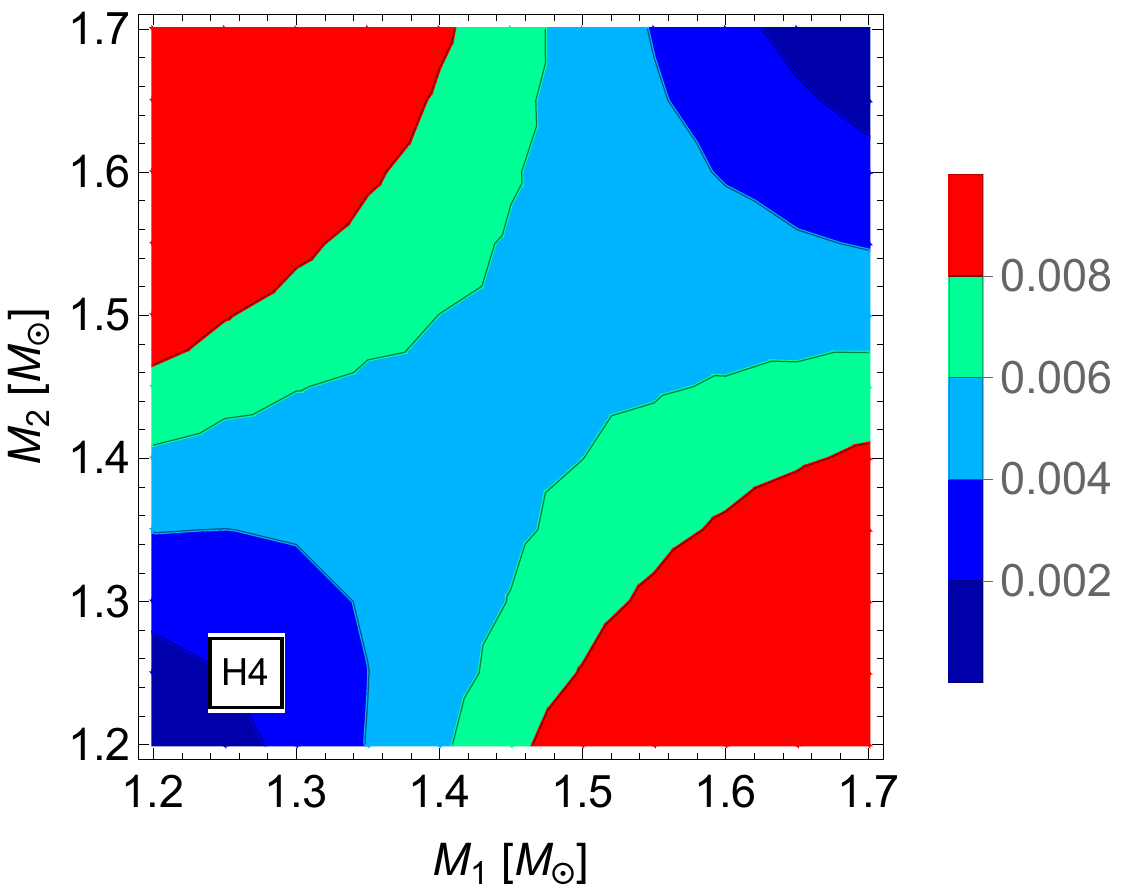}
\caption{Plots of the BNS mass ejected, depending on initial NS masses and EOS. For each, there is a symmetry about $Q=1$ because the NSs are indistinguishable. Dynamical mass ejected as a function of constituent NS masses is shown for the APR4 EOS (left-hand panel), ALF2 EOS (middle panel) and H4 EOS (right-hand panel).}
\label{fig:bnsmassejectedplot}
\end{figure*}

In Fig.~\ref{fig:bnsmassejectedplot}, we show the BNS dynamical mass ejected for APR4 EOS (left-hand panel), ALF2 EOS (our fiducial EOS, middle panel), and H4 EOS (right-hand panel) as a function of only the NS mass (the spin does not play a role). Since the NSs are indistinguishable except for their mass, the dynamical mass ejected is symmetric about a mass ratio $Q = 1$. The most dynamical mass ejection occurs when the system has the largest mass ratio and the least occurs when the mass ratio is 1. In general, for a stiffer EOS, more mass gets ejected. The BNS wind mass ejected depends only on the mass (and corresponding compactness) of the lighter NS, but monotonically decreases with increasing NS mass (see equation \ref{bnsdisk}). 

\begin{figure*}
\includegraphics[width=0.49\textwidth]{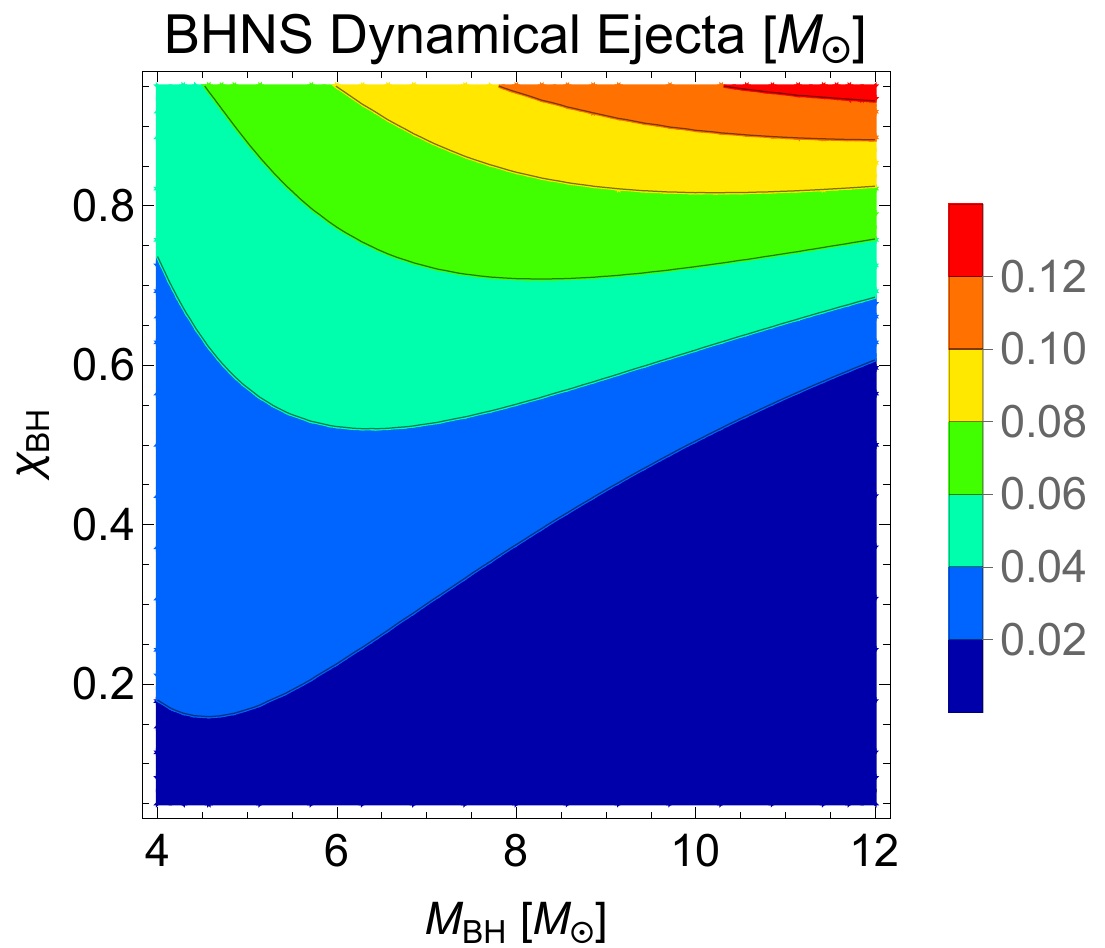}
\includegraphics[width=0.49\textwidth]{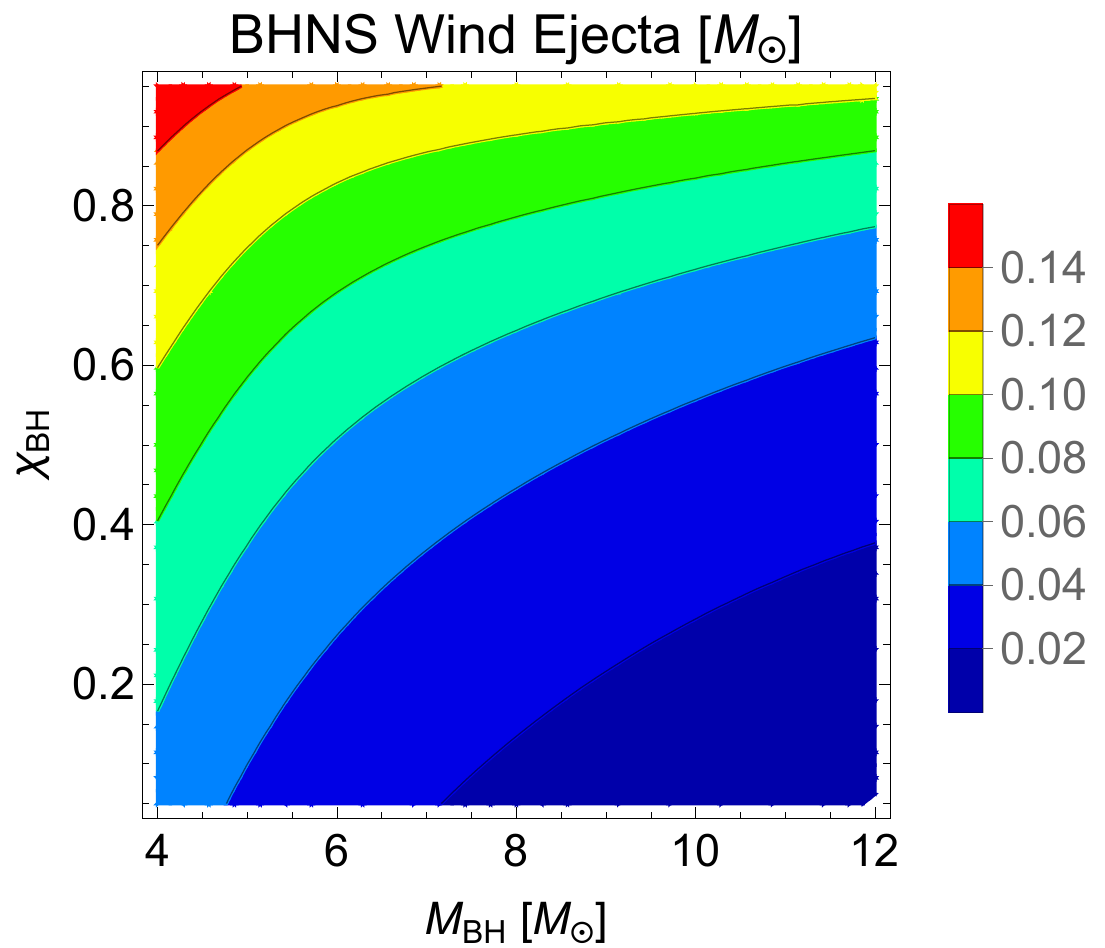}
\caption{Contour plots of the BHNS mass ejected for the ALF2 NS EOS, depending on initial $M_{\rm BH}$ and $\chi_{\rm BH}$, assuming a $1.3\,M_{\odot}$ NS. \textit{Left-hand panel:} dynamical mass ejected for BHNS mergers. The relationship between BH mass/spin and mass ejected is not monotonic here. \textit{Right-hand panel:} wind mass ejected for BHNS mergers. The relationship between BH mass/spin and mass ejected is monotonic in this case.}
\label{fig:bhnsmassejectedplot}
\end{figure*}

The relationship between mass ejected and initial compact object parameters differs in the BHNS case as well. In Fig.~\ref{fig:bhnsmassejectedplot} we show the dynamical (left-hand panel) and wind (right-hand panel) mass ejected as a function of $M_{\rm BH}$ and $\chi_{\rm BH}$, assuming a $1.3\,M_{\odot}$ NS. The least amount of mass ejection (and, in some cases, zero) occurs for high mass, low spin BHs. The mass ejected-BH parameter relationship is not monotonic for BHNS dynamical mass, however, and high mass, high spin black holes eject the most mass. This relationship is monotonic for BHNS wind mass where the low mass, high spin parameters eject the most wind mass. Like in BNS mergers, the mass ejected is higher for a stiffer NS EOS.

\label{lastpage}

\end{document}